\documentclass[a4paper, 11pt]{article}
\pdfoutput=1
\usepackage{jheppub}
\makeatletter
\def\@fpheader{}
\makeatother

\usepackage{graphicx,wrapfig,float,slashed}
\usepackage{bm}
\usepackage{amsmath,amssymb,dsfont,epsfig,graphicx}
\usepackage{mathtools} 
\usepackage{multirow,slashed}
\usepackage{blkarray}
\usepackage{epstopdf}
\usepackage{ragged2e}
\usepackage[utf8]{inputenc} 
\usepackage{cancel}
\usepackage[usenames,dvipsnames,table]{xcolor}
\usepackage[toc]{appendix}
\usepackage[normalem]{ulem}
\usepackage{enumitem}
\usepackage[font=small,skip=1pt]{caption}
\allowdisplaybreaks
\usepackage{pifont}
\usepackage{subcaption}
\usepackage{footnote}

\usepackage{feynmp-auto}
\newcommand{\feynwidth}{120}    
\newcommand{\feynheight}{80}    

\newcommand{\beq}{\begin{eqnarray}}
\newcommand{\eeq}{\end{eqnarray}}

\graphicspath{figures}

\newcommand{\su}{\widetilde{u}}
\newcommand{\sd}{\widetilde{d}}
\newcommand{\sq}{\widetilde{q}}

\newcommand{\Lum}{\mathcal{L}}
\newcommand{\sqbar}{\widetilde{q}^*}
\newcommand{\subar}{\widetilde{u}^*}
\newcommand{\sdbar}{\widetilde{d}^*}
\newcommand{\qcdp}{\text{QCD}^\prime}

\begin{document}

\title{Visible collider signals of natural quirks}
\author[1]{Joshua Forsyth,}
\emailAdd{jdfors21@byu.edu}
\author[2]{Matthew Low,}
\emailAdd{matthew.w.low@pitt.edu}
\author[1]{Carson Tenney,}
\emailAdd{catenney@student.byu.edu}
\author[1]{and Christopher B. Verhaaren}
\emailAdd{verhaaren@physics.byu.edu}
\affiliation[1]{Department of Physics and Astronomy, Brigham Young University, Provo, Utah, 84602, USA}
\affiliation[2]{Pittsburgh Particle Physics Astrophysics and Cosmology Center, 
Department of Physics and Astronomy, University of Pittsburgh, Pittsburgh, Pennsylvania, 15260, USA}

\abstract{
Though some LHC searches for new physics exceed the TeV scale, there may be discoveries waiting to be made at much lower masses. We outline a simple quirk model, motivated by models that address the hierarchy problem through neutral naturalness, in which new electroweakly charged states with masses as low as 100 GeV have not yet been probed by the LHC. We also describe a novel search strategy which is complementary to current search methods. In particular, we show its potential to discover natural quirks
over regions of parameter space that present methods will leave unexplored, even after the LHC's high-luminosity run.}

\preprint{}
\arxivnumber{}

\flushbottom
\maketitle
\flushbottom

\section{Introduction\label{sec:Intro}}

The Large Hadron Collider (LHC) continues to push the frontiers of our understanding of Nature's structure by producing ever greater luminosities and energies. The LHC's discovery of the Higgs boson~\cite{ATLAS:2012yve,CMS:2012qbp} and a host of quantum chromodynamics (QCD) bound states~\cite{LHCb-FIGURE-2021-001-report} demonstrate its power to pick out essential signals from the remains of proton collisions.  In addition, the LHC has made critical measurements of Standard Model (SM) parameters, such as masses and couplings. 

As important as these results are, discovering beyond the SM (BSM) dynamics is one of the primary drivers of the LHC mission.
Therefore, we must continually ask: are we utilizing the full discovery potential of the world's highest energy collider? In this article we discuss strategies for discovering motivated BSM states that are within the reach of the LHC's high-luminosity run. We show that novel search techniques can meaningfully enlarge the discovery power of the LHC. 

In particular, we focus on an extension of the SM which posits new particles charged under a confining gauge group, similar to the SM quarks. Unlike in QCD in the SM, however, it is assumed that all the states charged under this new gauge group have masses much larger than its confinement scale, $\Lambda'$. As a consequence, many of the properties of QCD dynamics, like the formation of jets, do not apply to these new fields.

Such an extension to the SM was first considered long ago~\cite{Okun:1979tgr,Okun:1980mu}, but was more recently characterized for the LHC era~\cite{Kang:2008ea}.  In particular, the latter work provided a thorough phenomenological investigation of these objects and coined the name ``quirks'' to refer to them.  
After production, a strongly coupled particle and antiparticle are connected by a tube of gauge flux as they fly apart.
In SM QCD this flux tube quickly fragments because it is energetically favorable to pair-produce quarks with masses well below $\Lambda_\text{QCD}$. For the case of quirks, in contrast, this pair-production rate is exponentially suppressed as there are no light states to pair-produce. Thus, the quirks are always bound to each other by this physical string of gauge flux. This can lead to long, potentially nonstandard tracks that are visible when the quirks are electrically charged~\cite{Knapen:2017kly,Farina:2017cts,Li:2020aoq,Sha:2024hzq}, as well as other interesting phenomena~\cite{Evans:2018jmd,Li:2019wce,Feng:2024zgp}.

While this notion of quirks is quite broad, here, we narrow our focus to the quirks that arise in theories motivated by the electroweak hierarchy problem~\cite{Craig:2022eqo}, specifically the framework that has come to be known as neutral naturalness~\cite{Batell:2022pzc}. These symmetry-based solutions to the hierarchy problem involve symmetry partners to the SM quarks that are not charged under SM QCD.  Typically, these quark partners are charged under an SU(3) gauge group with coupling related to the SM value at the scale of a few TeV. Because we have not yet discovered the fields charged under this gauge group, they are often referred to as a hidden sector.

Depending on the model these partner particles may be completely neutral under all SM gauge groups~\cite{Chacko:2005pe,Craig:2015pha,Cohen:2018mgv,Cheng:2018gvu} or they may carry SM electroweak charges~\cite{Burdman:2006tz,Cai:2008au}. The latter class of models is bounded by LEP searches to not have any electrically charged states below about 100 GeV~\cite{Egana-Ugrinovic:2018roi}. In these scenarios the new SU(3) gauge coupling runs to higher values faster than the SM QCD coupling, leading to confinement scales on the order of a few GeV~\cite{Curtin:2015fna}.\footnote{Variations, like a spontaneous breaking of the new color group~\cite{Batell:2020qad}, can modify this typical result.} This confinement scale ensures that quirk bound states appear point like compared to current collider tracking. 
However, as their dynamics, which differ significantly from the focus of most collider searches, are less-studied, the current set of searches are not suited to discover electroweakly charged quirks.

The defining feature of a quirk model, the gap between the mass of the states charged under the confining group and the confinement scale, implies that glueballs make up the lightest states of the bound state spectrum~\cite{Teper:1998kw,Morningstar:1999rf,Lucini:2010nv,Athenodorou:2021qvs}.
These works indicate that the lightest of these states have $J^{PC}$ quantum numbers of $0^{++}$, which means that it can mix with the SM Higgs and undergo long-lived decays to SM particles~\cite{Juknevich:2009gg}. In realizations of neutral naturalness this coupling arises (at minimum) from the top partners charged under the new color group. Consequently, several studies have considered how glueballs can be used to discover these frameworks~\cite{Craig:2015pha,Curtin:2015fna,Csaki:2015fba,Chacko:2015fbc}.

In this work we incorporate these glueballs into a new search strategy for these `natural quirks.' In doing so we strive to balance more general techniques and concrete models in which such quirk bound states appear~\cite{Burdman:2006tz,Cai:2008au,Curtin:2015jcv,Cheng:2018gvu,Xu:2018ofw,Cheng:2019yai,Ahmed:2020hiw}. Similar in spirit to works such as Refs.~\cite{Martin:2010kk,Li:2021tsy,Barela:2023exp,Li:2023jrt}, we employ a simplified model of scalar quirks (``squirks'') charged under a hidden SU(3) gauge group, which is motivated by folded supersymmetry (SUSY) constructions~\cite{Burdman:2006tz,Craig:2014fka,Cohen:2015gaa,Gherghetta:2016bcc}. While the results we find are specific to this model, many of the qualitative conclusions are expected to apply more generally, such as to other gauge groups and charge assignments.

A particularly visible signal of these squirks occurs when they make up an electroweak doublet. In this case bound states with a net electric charge of $Q=\pm 1$ (denoted $\Psi_\pm$) can be produced through a $W^\pm$. Due to their GeV confinement scale this bound state deexcites promptly and its constituents annihilate. Since $\Psi_\pm$ has electric charge, its dominant decay modes are to diboson final states: $ W^\pm\gamma$ or $W^\pm Z$. Both the ATLAS and CMS collaborations have searched for signals of new physics as $W\gamma$ resonances~\cite{ATLAS:2013way,ATLAS:2014lfk,ATLAS:2023kcu,CMS:2024ndg} and $WZ$ resonances~\cite{ATLAS:2019nat,ATLAS:2020fry,CMS:2021xor,ATLAS:2022zuc}.  Some of these searches have been recast as limits on electrically charged squirk bound states~\cite{Burdman:2008ek,Burdman:2014zta}. 

In addition, there have been phenomenological studies focused on how to strengthen the power of these searches.  Of particular note is Ref.~\cite{Capdevilla:2019zbx} which pointed out a tree-level zero in the differential cross section of the SM $W\gamma$ background. This observation might be utilized to reduce SM backgrounds and strengthen their discovery power but has not yet been implemented in LHC searches.

In this work, we reexamine this simple quirk system. We find that past works have not recognized that the diboson branching fractions are highly sensitive to the mass difference of the two states that make up the electroweak doublet. Mass splittings of even a few GeV lead to a significant weakening of the collider bounds. The upshot is that the LHC may be producing, but not discovering BSM states with electroweak charges with masses as low as one hundred GeV.  While the upcoming high-luminosity run will improve the discovery potential using these methods there is still a large region of motivated parameter space that will not be probed by the currently implemented search strategies.

To fully leverage this discovery opportunity, we describe a new search strategy for the high-luminosity run of the LHC. Rather than considering squirk production only, we investigate the production of the squirk bound state along with a gluon of the new confining gauge group.  This hidden sector gluon must hadronize into a hidden sector glueball, which often produces a displaced vertex.  While the production cross section for this process is lower than squirk pair production alone, the striking combination of a prompt diboson resonance (from the squirk annihilation) and the displaced vertex (from the glueball) is expected to be effectively background-free. Therefore, this strategy allows for a powerful new probe of many squirk scenarios left undiscovered by current practices. 

In the following section, Sec.~\ref{s.quirks}, we review the dynamics of quirk bound states and the glueballs associated with their confining gauge group.  We estimate the time it takes these bound states to deexcite due to photon or hidden sector glueball radiation and review how glueballs of the new sector can decay into SM states. Sec.~\ref{s.model} defines a concrete, simple model that is used to demonstrate our new search strategy. The current limits on the model parameters are determined as well as the projected limits for the LHC's high-luminosity run. In Sec.~\ref{s.collider} we describe a new search strategy, which utilizes the production of a hidden sector glueball along with an electrically charged squirk bound state. Our results show that these methods are complementary to existing techniques and can be used to probe regions of parameter space that might otherwise be left unexplored by the LHC. We conclude in Sec.~\ref{sec:Con} and comment on interesting directions for further work.

\section{Review of Quirks and Glueballs \label{s.quirks}}

This section reviews the dynamics of quirk bound states from production, through deexcitation, and to annihilation.  This spin-independent, semiclassical analysis applies to both squirks and quirks. In the rest of this work we use the term quirk generally when results apply to both fermionic and scalar particles. When the distinction is important we consistently use the term ``squirk'' to refer specifically to scalar quirks. For instance, several particulars of the bound state analysis are specific to the scalar nature of the particles we focus on in this work. We also review some aspects of glueballs in the pure Yang-Mills regime, especially how the lightest hidden sector glueball can decay into SM states through the Higgs.

\subsection{Quirk Dynamics}

At low energies, states charged under a confining group may be described as experiencing a linear potential
\beq
V(r)\approx\sigma r, \label{e.V}
\eeq
where the so-called string tension $\sigma$ is related to the scale of confinement, $\Lambda'$, by $\sigma\approx3.6\Lambda^{\prime \,2}$~\cite{Lucini:2004my,Teper:2009uf} and $r$ is the distance between particle and antiparticle. This parameter plays a significant role when determining the dynamics of quirk bound states.

At a high energy collider, like the LHC, a pair of particles (with masses $m_1$ and $m_2$) are produced at essentially zero separation.\footnote{If there is nonzero separation then there is some amount of energy stored in the flux tube to begin with, but it is small compared to the kinetic energy $E_k$ if the quirks are not produced near threshold so the analysis above still holds.} In the center of momentum (CM) frame, the magnitude of the initial three-momentum of each particle is
\beq
    |\vec{p}|\equiv p(0)=\frac{E_T}{2}\sqrt{1-\frac{M^2+\Delta^2}{E_T^2}+\frac{M^2\Delta^2}{E_T^4}}~,
\eeq
where $E_T$ is the total energy of the produced particles, $M=m_1+m_2$ is the total mass, and $\Delta=m_1-m_2$ is the mass splitting.

As the particles propagate away from each other the potential between them can be modeled by Eq.~\eqref{e.V}.
In the SM, the string connecting the particles is fragmented by the production of pairs of light quarks. For a pair of quirks, however, the string remains unbroken
and determines the subsequent dynamics. The particles in the bound state evolve in time according to Newton's second law,
\beq
\frac{dp}{dt}=-\sigma. \label{e.N2}
\eeq
It follows that the velocity is 
\beq
    v(t)=\frac{p(0)-\sigma t}{\sqrt{m_q^2+\left( p(0)-\sigma t\right)^2}}~,
\eeq
where $m_q$ is the mass of the particle. This shows that the bound quirks move apart with decreasing velocity until the turning time $t_\text{turn}$ when the velocity becomes zero.  The turning time is
\beq
    t_\text{turn}=\frac{p(0)}{\sigma}.
\eeq
The particles then start to approach one another, and assuming they do not annihilate, begin an oscillating trajectory.

A full period of oscillation starts from creation, then to the maximum separation between quirks where they have zero velocity, then back past the creation point. They then continue the other way to another turning point and then finally back to the initial configuration with the quirks moving apart. If we assume negligible energy loss, the period of one full oscillation is simply $T=4t_\text{turn}$.  In reality, the quirks can radiate energy into any boson they couple to. This rate is typically small over a single period~\cite{Harnik:2008ax} so this estimate for the period should be approximately correct.

If we take into account that energy is lost during every period, we can estimate the time of 
a single
oscillation by
\beq
    T=\frac{4p}{\sigma},
\eeq
where $p$ now represents the momentum of a quirk in the CM frame at the beginning of that period. The angular frequency is
\beq \label{e.omega}
    \omega=\frac{\pi\sigma}{2p}
    = 1.8\frac{\Lambda^{\prime\,2}}{p}
    \approx 283~{\rm MeV} \left(\frac{\Lambda^{\prime}}{5~{\rm GeV}}\right)^2\left(\frac{500~{\rm GeV}}{p}\right).
\eeq
Neglecting the radiation of hidden sector gluons for a moment, the bound state loses energy through the radiation of photons as the quirks oscillate. The distribution of photon energies is sharply peaked at the characteristic frequency of the system given in Eq.~\eqref{e.omega}~\cite{Harnik:2008ax}. This confirms that the emitted photons are quite soft when the quirk kinetic energy is large. Indeed, the bound state radiates much of its energy away into these soft photons, the distribution of which has been identified as a way to search for quirks~\cite{Harnik:2008ax}. 
 
A more precise calculation of the power radiated into photons is obtained from the force from Eq. \eqref{e.N2}. This implies that the magnitude of the relativistic acceleration is 
\beq
a(t)=\frac{\sigma}{m_q}\left( 1-v^2 \right)^{3/2}~,
\eeq
which we insert into the relativistic Larmor formula. The radiated power into massless gauge bosons by each quirk is
\begin{align}
    \mathcal{P}_q=&\frac{8\pi\alpha}{3(1-v^2)^3}\left[a^2(1 -v^2)+(\vec{v}\cdot \vec{a})^2\right]
    = \frac{8\pi\alpha\sigma^2}{3m_q^2},
\end{align}
where we have assumed the velocity and acceleration are always either parallel or antiparallel. By integrating the total power radiated by each quirk, $\mathcal{P}=\mathcal{P}_{q_1}+\mathcal{P}_{q_2}$, over the full period we find the energy radiated is
\begin{align}
E_\text{period}=&\frac{32\pi\alpha\sigma p}{3}
    \left( \frac{1}{m_1^2}+\frac{1}{m_2^2} \right)
    =\frac{256\pi\alpha\sigma p}{3M^2}\frac{1+\frac{\Delta^2}{M^2}}
    {\left(1-\frac{\Delta^2}{M^2}\right)^2}~.
\end{align}
The time to deexcite to the ground state ($E_T-M\sim 0$) is
\beq
    t_\text{deexcite}\sim\frac{E_T-M}{\mathcal{P}}
    =\frac{3M^3}{64\pi\alpha\sigma^2}
    \frac{\left(\frac{E_T}{M}-1\right)
    \left(1-\frac{\Delta^2}{M^2}\right)^2}{1+\frac{\Delta^2}{M^2}}~. \label{e.t_dE}
\eeq
This is the maximum time it takes the bound state to deexcite. Any additional radiation\textemdash such as hidden glueballs\textemdash can only speed up the process. 

One might hope that radiation into hidden sector gluons is similar to photon radiation. There are, however, several important distinctions. While a single photon can be radiated with frequency set by the oscillations of the system, a single hidden sector gluon cannot be radiated as it is not a color singlet. At least two must be emitted to form a glueball and the mass of the lightest glueball state scales like $m_0\gtrsim \Lambda^{\prime}$~\cite{Lucini:2008vi,Lucini:2010nv,Athenodorou:2021qvs}. Thus, we might expect the glueball radiation to go not like $\alpha_s$, but $\alpha_s^2$ and to be suppressed by $\omega/m_0$ in some way. This suppression may be exponential, considering the photon distribution's sharp peak at the angular frequency of oscillation~\cite{Harnik:2008ax}.

As perturbative radiation of hidden sector gluons looks unlikely, one might turn to nonperturbative processes. If the quirks pass by one another within a radius of $\Lambda^{\prime\,-1}$ they enter a region of space where nonperturbative interactions dominate. Here there is essentially a large hidden sector gluon background, and hence no penalty to radiating multiple hidden sector gluons. The relevant scales are $\Lambda^{\prime}$ and the glueball mass, $m_0$, so the classical oscillation frequency is not expected to matter. Of course, making a reliable prediction of this process is difficult. Radiation of a hard hidden sector gluon in this scenario has an expected amplitude scaling of $m_0^{-6}$ for fermionic quirks~\cite{Kang:2008ea} or $m_0^{-8}$ for squirks~\cite{Cheng:2018gvu}. If the gluon has sufficient energy, it can pull a second gluon from the vacuum to form a glueball. While this argument has been employed in some studies, the reliability of this approach is unclear. 

In what follows we simply assume that glueballs are not radiated in the deexcitation process. If they are, deexcitation becomes quicker which may increase the signal in the searches we outline below. Therefore, our analysis is a conservative one. This assumption also implies that we cannot confidently use softly radiated glueballs to discover quirk bound states. If a reliable method of estimating glueball radiation is obtained, exploring how such radiated glueballs can be used to discover quirks would be interesting.

While we do not consider glueball radiation from deexcitation, glueballs do play an essential role in our work. These are produced as a part of the hard scattering where perturbative methods are well understood. Therefore, we review some important aspects of glueballs below.

\subsection{Glueball Dynamics\label{ss:glueDyn}}

Though hidden sector glueballs radiated through deexcitation play no role in what follows, these lightest bound states of the hidden confining sector are still essential to our work.

The glueball spectrum for SU(3) has been determined in lattice QCD~\cite{Teper:1998kw}. The lightest state, $J^{PC}=0^{++}$, has mass about $m_0\approx6.8\Lambda^{\prime}$~\cite{Morningstar:1999rf}. For confinement scales of a GeV, as typically arise in realizations of neutral naturalness, this implies the lightest glueball has a mass of a few tens of GeV. Similar results hold for general SU($N$), always with $0^{++}$ as the lightest glueball state~\cite{Lucini:2008vi,Lucini:2010nv,Athenodorou:2021qvs}. Hence, while we focus on the SU(3) case, a similar analysis applies to a wide variety confining hidden sectors.

The quantum numbers of the lightest glueball allow it to mix with the visible sector Higgs. The strength of the mixing depends on the structure of the hidden and visible sectors, but can be parametrized as~\cite{Juknevich:2009gg}
\begin{align}
    \mathcal{L} &=
    \frac{\alpha'}{3\pi}\left(\frac{y^2}{\mu^2}\right) |H|^2 G^{(a)}_{\mu\nu}G^{(a)\mu\nu},
    \nonumber\\
    &=
    \frac{\alpha'}{3\pi}\left(\frac{y^2}{\mu^2}\right) vh G^{(a)}_{\mu\nu}G^{(a)\mu\nu}+\mathcal{O}(h^2)~,
\end{align}
where $h$ is the 
physical Higgs boson after it gets a vacuum expectation value $v/\sqrt{2}$ and the tensor $G^{(a)}_{\mu\nu}$ is the hidden sector gluon field with coupling $\alpha'$. The combination $(y/\mu)^2$ depends on the mass ratio of the SM top-quark to hidden sector states with large coupling to the Higgs field. In the case of folded SUSY, for example, this coupling is generated by the folded squark loops that couple the Higgs to the hidden gluons. 

As shown in Ref.~\cite{Curtin:2015fna}, the leading value for this coupling in folded SUSY is
\beq
    \frac{y^2}{\mu^2} = \frac{m_t^2}{16v^2}
    \left( \frac{1}{m_{\tilde{t}_1}^2}+\frac{1}{m_{\tilde{t}_2}^2}
    -\frac{X_{\tilde{t}}}{m_{\tilde{t}_1}^2 m_{\tilde{t}_2}^2} \right),
\eeq
where $X_{\tilde{t}}$ denotes the mass mixing in the stop sector and $m_{\tilde{t}_i}$ are the two stop eigenstates. It is useful for us to define a dimensionless parameter $c_g$ such that
\beq
    \frac{y^2}{\mu^2} = \frac{c_g}{4v^2}.
\eeq
For folded stops of twice the top-quark mass and $X_{\tilde{t}}=0$ we find $c_g=1/8$, which we take as a useful upper bound. As either the coupling of these particles to the Higgs is reduced or the mass is increased the value of $c_g$ goes down. Rather relying on a specific model, however, we simply use this parameter\textemdash along with the lightest glueball mass $m_0$\textemdash to characterize the dynamics of hidden glueballs.

As mentioned above, the lightest glueball can decay into SM states by mixing with the Higgs~\cite{Juknevich:2009gg}. The $0^{++}$ decay width is given by
\beq
    \Gamma(0^{++}\to X_\text{SM}X_\text{SM})=|c_g|^2\left( \frac{\alpha'}{6\pi}\frac{f_{0^{++}}}{v(m_h^2-m_0^2)}\right)^2\Gamma\left(h(m_0)\to X_\text{SM}X_\text{SM}\right)~, \label{e.decay-m0}
\eeq
where $m_h=125$ GeV is the mass of the Higgs boson and $\Gamma(h(m_0)\to X_\text{SM}X_\text{SM})$ is the decay width of a SM Higgs of mass $m_0$. The parameter $f_{0^{++}}=\langle0|\text{Tr}\,G_{\mu\nu}G^{\mu\nu}|0^{++}\rangle$ has been shown by lattice calculations to be proportional to the cube of the lightest glueball mass $4\pi\alpha'f_{0^{++}}\propto m_0^3$~\cite{Chen:2005mg,Meyer:2008tr}. Reference~\cite{Chen:2005mg} found that $4\pi\alpha' f_{0^{++}}/m_0^3=3.14\pm 0.83$  while Ref.~\cite{Meyer:2008tr} obtained $4\pi\alpha' f_{0^{++}}/m_0^3 =2.69\pm 0.27$. In our analysis we use the second, more recent and reportedly less uncertain, value. See Appendix~\ref{app.f0} for further discussion of how these values are obtained from the lattice results.

\begin{figure}[th]
    \centering
    \includegraphics[width=0.8\linewidth]{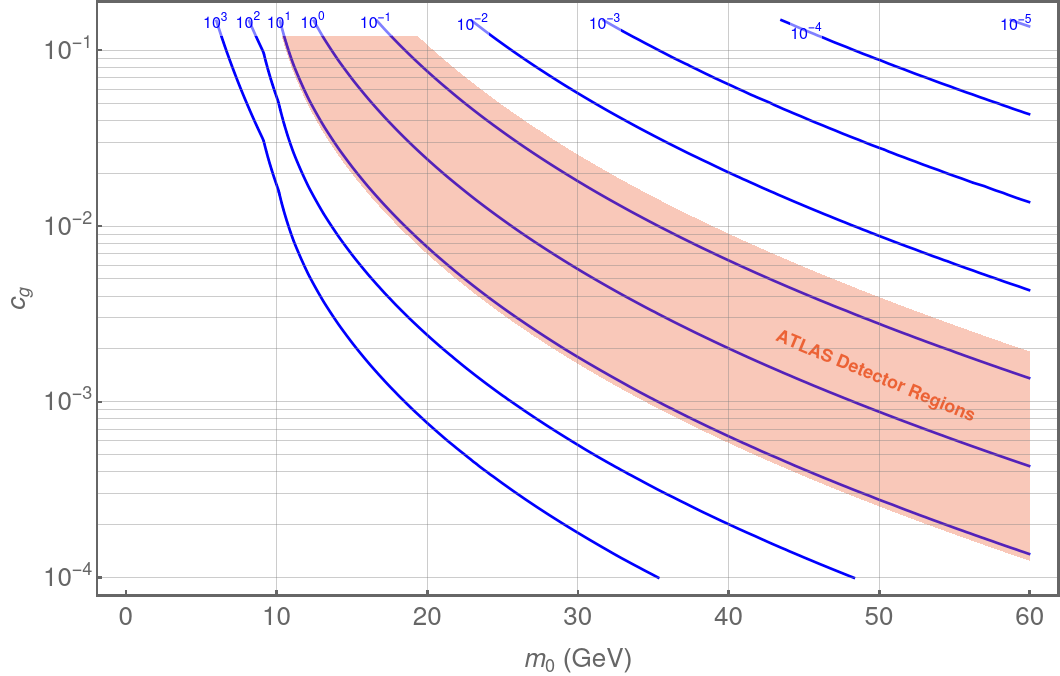}
    \caption{Decay length in meters of the lightest glueball state, $J^{PC}=0^{++}$. This decay length is plotted as a function of the glueball mass, $m_0$, and the coupling parameter, $c_g$, which characterizes the coupling between the hidden gluon field and the SM Higgs. The red region indicates ATLAS detector regions sensitive to displaced vertices.}
    \label{fig:glueballDecayLength}
\end{figure}

As the glueball decay width is proportional to the decay width of a SM Higgs of the glueball mass, we use HDECAY 6.61~\cite{Djouadi:1997yw,Djouadi:2018xqq} to determine the Higgs width as a function of mass. Using one-loop running (see Appendix~\ref{apss:runalpha} for details) we find the value of $\alpha'$ at the lightest glueball mass is
\beq
    \alpha^\prime(m_0)\approx 0.21~. \label{e.alpha-m0}
\eeq
Using this in Eq.~\eqref{e.decay-m0} we find the decay widths for $0^{++}$ glueballs of various masses and couplings. 

In Fig. \ref{fig:glueballDecayLength} we plot the glueball decay length as a function of the lightest glueball mass and coupling of the hidden gluons to the Higgs. We also highlight, in red, the region from 5 cm to 12 m, which approximates the regions of the ATLAS detector used to detect displaced particle decays. This makes clear that over a range of values there are opportunities to use glueball decays to reduce SM backgrounds when searching for quirks. In Sec.~\ref{s.collider}, we show how the production of a glueball along with a quirk bound state can significantly enhance experimental searches for such states. In the following section we define the model used to show the power of these searches.

\section{Simplified Scalar Quirk Model}\label{s.model}

For this work we use a simple extension to the SM that includes one generation of squirks (scalar quirks) with gauge charges given in Table \ref{t.quantumNumbers}. These states resemble a generation of folded squirks that arise in folded SUSY. Importantly, they are neutral under SM color, but are charged under a different, confining SU(3) gauge group.
\begin{table}[h]
\centering
\begin{tabular}{c|cccc}
    & $\text{SU}(3)_{\qcdp}$ & $\text{SU}(3)_c$ & $\text{SU}(2)_L$ & $\text{U}(1)_Y$ \\
    \hline
    $\widetilde{Q}$ & {\bf 3} & {\bf 1} & {\bf 2} & $1/3$ \\
    $\widetilde{U}$ & $\mathbf{\bar{3}}$ & {\bf 1} & {\bf 1} & $-4/3$ \\
    $\widetilde{D}$ & $\mathbf{\bar{3}}$ & {\bf 1} & {\bf 1} & $2/3$ \\
\end{tabular}
\caption{Quantum numbers for the SM-color-neutral squirks in our model.
\label{t.quantumNumbers}}
\end{table}

The up-type squirk (labeled as $\su$) has a mass mixing matrix
\beq
    m_{\su}^2 = 
    \begin{pmatrix}
        m_{\widetilde{Q}_1}^2 + m_u^2 + D_{\su_L} & m_u X_u \\
        m_u X_u^* & m_{\widetilde{U}_1}^2 + m_u^2 + D_{\su_R}
    \end{pmatrix}~,
\eeq
where $m_i$ are masses of the respective fields, $X_u$ is a mixing term that comes from interactions with the Higgs, and the $D$-terms are given by
\begin{align}
    D_{\su_L} &= 
    \left(\frac{1}{2}-\frac{2}{3}s_W^2\right)m_Z^2\cos{2\beta}~,& 
    D_{\su_R} &= 
    \frac{2}{3}m_Z^2s_W^2\cos{2\beta}~,
\end{align}
with $s_W\equiv\sin\theta_W$ which is the sine of the weak mixing angle. The parameter $\beta$ is the standard parameter for describing the ratio of the vacuum expectation values of the two Higgs doublets in supersymmetric theories, with $\tan{\beta}>1$. We can construct a similar mass matrix for the down-type squirk (labeled as $\sd$) by replacing the $u$ and $U$ subscripts with $d$ and $D$ and using the $D$-terms
\begin{align}
    D_{\sd_L} &= 
    -\left(\frac{1}{2}-\frac{1}{3}s_W^2\right)m_Z^2\cos{2\beta}~,&
    D_{\sd_R} &= 
    \frac{1}{3}m_Z^2s_W^2\cos{2\beta}~.
\end{align}
To focus on the aspects of this construction most relevant to the novel search strategy we are describing, we consider the simple case of $\su_R$ and $\sd_R$ being lifted out of the spectrum, leaving just the doublet of $\su_L$ and $\sd_L$. That is, we take $m_{\widetilde{U}_1,\widetilde{D}_1}\gg m_{\widetilde{Q}_1} \gg m_{u,d}$. If we neglect the off-diagonal terms we find that the difference in masses leads us to a mass splitting
\beq \label{e.Delta}
    \Delta\equiv m_{\su}-m_{\sd} = \frac{m_{\su}^2-m_{\sd}^2}{m_{\su}+m_{\sd}} = \frac{m_W^2\cos{2\beta}}{M}~,
\eeq
where $m_W$ is the mass of the $W$ and $M=m_{\su}+m_{\sd}$ is the combined mass of the squirks.

For $\tan{\beta}\approx 10$ and $200~{\rm GeV}< M<2~{\rm TeV}$, we find an expected splitting in the range $-32~{\rm GeV}\lesssim\Delta\lesssim -3~{\rm GeV}$. In general, therefore, we expect a mass splitting between squirk flavors of a few to tens of GeV, though nonzero $X_{u,d}$ terms affect the tight relation given above.  In practice, we simply take $M$ and $\Delta$ as free parameters to illustrate which mass configurations our strategy is most sensitive to. 

These states can be produced at hadron colliders at tree-level through $s$-channel SM vector bosons, see Fig.~\ref{fig:production}. The analytic expression for the parton-level cross section is given in Appendix~\ref{app.production}. A bound state produced through a $W$ (for example one having constituents $\sq_u$ and $\sqbar_d$), has net SM electric charge and is labeled by $\Psi_\pm$. Production through a $\gamma$ or $Z$ leads to charge neutral bound states, which we denote by $\Psi_0$.

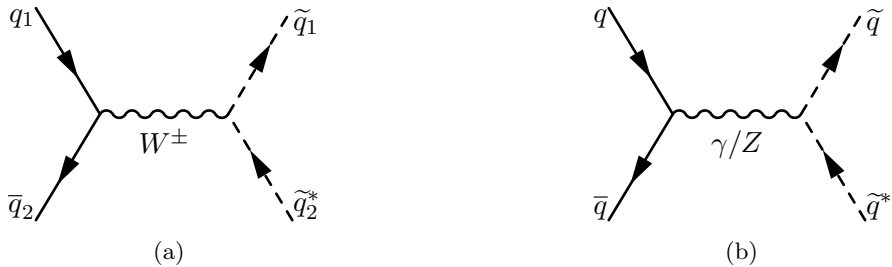
\begin{figure}[bh]
    \begin{subfigure}{0.49\textwidth}
    \centering
    \begin{fmffile}{production_charged}
        \begin{fmfgraph*}(\feynwidth,\feynheight)
        \fmfleft{i1,i2}
        \fmfright{o1,o2}
        \fmf{fermion}{i2,v1,i1}
        \fmf{boson,label=$W^\pm$}{v1,v2}
        \fmf{scalar}{o1,v2,o2}
        
        \fmfv{label=$\overline{q}_2$,label.dist=0.5,label.angle=135}{i1}
        \fmfv{label=$q_1$,label.dist=0.5,label.angle=-135}{i2}
        \fmfv{label=$\sqbar_2$,label.dist=0.5,label.angle=45}{o1}
        \fmfv{label=$\sq_1$,label.dist=0.5,label.angle=-45}{o2}
        \end{fmfgraph*}
    \end{fmffile}
    \caption[chargedProduction]{}
    \end{subfigure}
    \begin{subfigure}{0.49\textwidth}
    \centering
    \begin{fmffile}{production_neutral}
        \begin{fmfgraph*}(\feynwidth,\feynheight)
        \fmfleft{i1,i2}
        \fmfright{o1,o2}
        \fmf{fermion}{i2,v1,i1}
        \fmf{boson,label=$\gamma/Z$}{v1,v2}
        \fmf{scalar}{o1,v2,o2}
        
        \fmfv{label=$\overline{q}$,label.dist=0.5,label.angle=135}{i1}
        \fmfv{label=$q$,label.dist=0.5,label.angle=-135}{i2}
        \fmfv{label=$\sqbar$,label.dist=0.5,label.angle=45}{o1}
        \fmfv{label=$\sq$,label.dist=0.5,label.angle=-45}{o2}
        \end{fmfgraph*}
    \end{fmffile}
    \caption[neutralProduction]{}
    \end{subfigure}
  \caption{Tree-level production of squirk bound states from SM quarks. Diagram (a) produces a bound state $\Psi_\pm$ with net electric charge, while the bound state $\Psi_0$ in (b) is electrically neutral.}
  \label{fig:production}
\end{figure}

If there is a mass splitting between the up-type and down-type squirks, then the heavier state can $\beta$-decay into the lighter. This can change a charged bound state into a neutral one, or a neutral state composed of two heavy squirks into neutral state composed of the two light squirks. This is particularly relevant to $\Psi_\pm$ decays, which are the focus of this paper.

If a charged bound state $\beta$-decays into a light neutral state before it fully deexcites, then the search strategies outlined below do not apply. Therefore, determining the timescale for $\beta$-decay relative to deexcitation is essential. The three-body decay width of the heavy squirk through an off-shell $W$ is given explicitly in Appendix \ref{app.beta}, but roughly scales like
\beq \label{e.t_beta}
    \tau_\beta \sim \left(10^{-12}~\text{s}\right)
    \left(\frac{10~\text{GeV}}{\Delta}\right)^5~,
\eeq 
where $\Delta$ is the mass difference between the squirks.
The probability for a bound state to fully deexcite before $\beta$-decay is then
\beq \label{e.P_DE}
    P_\text{deexcite} = e^{-\frac {t_\text{deexcite}}{\tau_\beta}}~,
\eeq
where $t_\text{deexcite}$ is given in Eq.~\eqref{e.t_dE}. 

A bound state with $M=1000$ GeV and $\Delta=10$ GeV produced at $E_T\approx$ 2 TeV has a $99.9\%$ probability to fully deexcite before $\beta$-decay. This probability drops, however, to $97.5\%$ at $\Delta=20$ GeV, to $82.1\%$ at $\Delta=30$ GeV, and to $42.4\%$ at $\Delta=40$ GeV. The drop in probability is steeper for larger $M$ values, because less of the energy is available for motion, which drives radiation and deexcitation. Thus, while we do consider splittings as large as $\Delta\approx 40$ GeV, as $\Delta$ and $M$ increase the signal strength suffers, independent of production cross section. 

In addition to the time a bound state takes to deexcite, we must consider the characteristic time for the particles in the ground state to annihilate: 
\beq
t_\text{ann}=\Gamma_\text{T}^{-1}~,
\eeq
where $\Gamma_\text{T}$ is total decay width of the ground state.
Using the formulas for the decay widths given in Appendix~\ref{app.widths} we find that the lightest states with the smallest mass splitting and smallest hidden sector glueball mass have decay times on the order of $10^{-19}$ seconds. Increasing the total mass, mass splitting, or hidden sector glueball mass reduces this time further. As these timescales are orders of magnitude smaller than those of deexcitation or $\beta$-decay, in our simulations we treat annihilation as instantaneous.

\subsection{Indirect Constraints}\label{ss.model_constraints}

A strong indirect limit on our squirk model comes from the mass splitting between the two states. This difference in masses contributes to the $T$ parameter~\cite{Peskin:1991sw}, or equivalently to the $\rho$ parameter~\cite{ParticleDataGroup:2024cfk}. We parametrize $\rho_0$ as
\begin{equation}
\rho_0 = \frac{m_W^2}{m_Z^2 c_W^2 \hat{\rho}},
\end{equation}
where $c_W\equiv\cos\theta_W$ and $\hat{\rho}$ incorporates the SU(2)-breaking effects from the Standard Model such that for the Standard Model alone, $\rho_0 = 1$.  The measured value is $\rho_0 = 1.00031 \pm 0.00019$~\cite{ParticleDataGroup:2024cfk}.

The contribution of a squirk doublet, with masses $m_1$ and $m_2$, to $\rho_0$ is
\begin{align}
\Delta \rho_0 &= 
\frac{N_c' G_F}{8\sqrt{2}\pi^2} \left(m_1^2 + m_2^2
- \frac{4 m_1^2 m_2^2}{m_1^2 - m_2^2} \ln \frac{m_1}{m_2} \right)\\
&= \frac{N_c' G_F M^2}{16\sqrt{2}\pi^2} \left[1 + \frac{\Delta^2}{M^2} -\left(1-\frac{\Delta^2}{M^2}\right)^2\frac{\tanh^{-1}\frac{\Delta}{M}}{\frac{\Delta}{M}}\right]~,
\end{align}
where $G_F$ is the Fermi constant and $N'_c$ is the number of colors in the new confining gauge group (three for our model). For $M>100$~GeV the limit on the splitting, $\Delta=m_1-m_2$, is nearly constant.  At 95\% confidence level this limit is $4.9~{\rm GeV} < \Delta < 43.0~{\rm GeV}$. There is a lower limit because the existing measurement is more than $1\sigma$ from the SM prediction.  We conservatively neglect the lower limit which allows for additional new physics contributions in addition to the squirks and take $\Delta < 43.0~{\rm GeV}$ as the robust indirect limit on the mass splitting. We saw above that mass splittings of more than about 40 GeV lead to fast $\beta$-decay times, so this result does not impose much additional constraint upon the parameter space probed by our search strategies.

\subsection{Direct Constraints}\label{ss.decay_states}

In this simple model, both the charged and neutral bound states can decay into SM particles. The dominant decay mode of the neutral squirk bound states is into hidden sector gluons, which shower and hadronize into hidden sector glueballs.  As discussed in Sec.~\ref{ss:glueDyn}, the lightest of these hidden sector glueballs may decay, through mixing with the Higgs, into SM states. This has been discussed as a potential discovery channel for these states~\cite{Chacko:2015fbc}. The modeling of pure gluon parton showering and hadronization has recently taken significant steps forward~\cite{Curtin:2022tou,Batz:2023zef}. However, the careful treatment of these neutral bound state decays\textemdash especially those that decay through hidden sector glueballs\textemdash is not considered here, but will be explored in upcoming work. In this article, we confine our analysis to signatures that include a prompt SM resonance from the decay of the squirk bound state.

\subsubsection{Bounds on Neutral Squirk Bound States}

As described in Ref.~\cite{Kang:2008ea}, the squirks are unlikely to annihilate until they have shed sufficient energy to reach their ground state. Consequently, the annihilation products of zero-angular-momentum bound states dominate the visible signals produced by these states. The lowest angular momentum decay modes of neutral squirk bound states $\Psi_0$ are into pairs of gauge bosons: $\gamma\gamma$, $\gamma Z$, $ZZ$, and $WW$. If the states have significant coupling to the Higgs then additional Higgs processes play a role~\cite{Barela:2023exp}, but we assume a sufficiently small Higgs coupling to the squirks such that these can be neglected. 

\begin{figure}[th]
    \centering
    \begin{subfigure}{0.49\textwidth}
        \includegraphics[width=\textwidth]{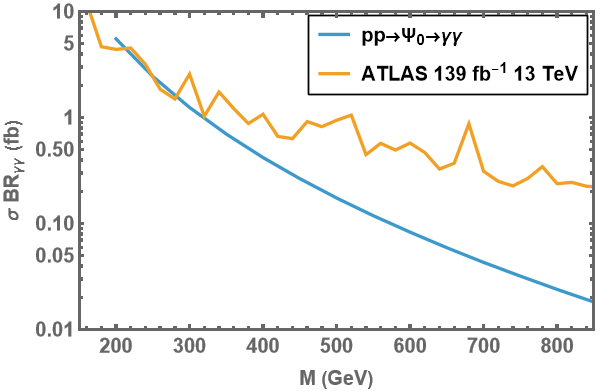}
        \caption[Image 1]{\label{fig:bounds_gamgam}}
    \end{subfigure}
    \begin{subfigure}{0.49\textwidth}
        \includegraphics[width=\textwidth]{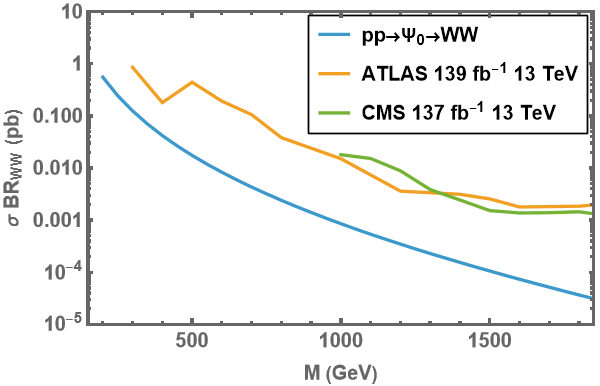}
        \caption[Image 2]{\label{fig:bounds_WW}}
    \end{subfigure}
    \caption{Potential collider bounds from SM final states of decays of the neutral squirk bound state $\Psi_0$. On the left (right) the bound is from decays into $\gamma\gamma$ ($WW$), assuming a conservative branching fraction of 1\% (100\%). We see that there is a lower bound on the combined squirk mass $M$ from diphoton searches.}
    \label{fig:NeutBounds}
\end{figure}

In terms of collider signals, the diphoton channel is the cleanest and most likely to lead to bounds from existing collider studies.  The calculation of the various neutral state decay widths shows that the branching fraction into two photons is never more than about 1\%. This allows us to make a conservative estimate of the lower bound. We simulate the electroweak production of these states (using up-type quark electroweak charges) in MadGraph~\cite{Alwall:2014hca} and then apply a 1\% branching fraction and compare with the bounds from Ref.~\cite{ATLAS:2021uiz}. The results are shown in the left panel of Fig.~\ref{fig:NeutBounds}. We find the lower bound on the total mass $M$ of the bound state to be about 200~GeV.

When the squirk mass splitting $\Delta$ is closer to 40~GeV, the branching fraction into $WW$ final states is significantly enhanced for TeV-scale squirk masses. To account for this we also compare the squirk production cross section to the $WW$ resonance bounds from ATLAS~\cite{ATLAS:2020fry} and CMS~\cite{CMS:2021klu}. In effect, we are taking the very conservative overestimate that the branching fraction is 100\%. Even with this choice we find, as shown in the right panel of Fig.~\ref{fig:NeutBounds}, that the cross section to produce a $\Psi_0$ bound state that decays into $WW$ is well below the experimental bound. Hence, there is no additional constraint on the model from this channel. All other decays into SM states are similarly unconstrained.

\subsubsection{Bounds on Charged Squirk Bound States\label{sssec:ChargedBounds}}

Collider signals from charged squirk bound states $\Psi_\pm$ are larger than those from neutral states because charge conservation prevents them from decaying into hidden sector gluons. The only open decays for the ground state $\Psi_\pm$ are into $W\gamma$ and $WZ$. Decay through an $s$-channel $W$ into SM fermions is forbidden for the ground state. The reason is the vanishing relative velocity of the ground state quirks~\cite{Burdman:2008ek}. 

As with the neutral bound states, the production and annihilation of the charged states can be compared to LHC searches. Previous studies~\cite{Burdman:2008ek,Burdman:2014zta} applied these methods only when the squirks were degenerate. In the $\Delta=0$ limit, the branching fraction to $W\gamma$ dominates. As this final state is experimentally simpler than $WZ$, the experimental reach is correspondingly greater.

A careful analysis of the branching fractions, however, shows that the $\Delta=0$ results change rapidly away from that limit. We find that the ratio of the decay widths is
\begin{align}
\frac{\Gamma_{WZ}}{\Gamma_{W\gamma}}=&\frac{s^2_W\beta_{WZ}}{c^2_W\beta_{W\gamma}}
\left[1\phantom{\frac{M^2\Delta^2}{8m_W^2m_Z^2s^4_W(Q_u+Q_d)^2}}\right.\label{eq:widthRatio}\\
&\left.+\frac{M^2\Delta^2}{8m_W^2m_Z^2s^4_W(Q_u+Q_d)^2}\left(1+\frac{2\Delta}{M(Q_u+Q_d)}+\frac{3\Delta^2}{M^2(Q_u+Q_d)^2}-2s^2_W\frac{m_Z^2}{M^2} \right) \right] +\mathcal{O}\left(\frac{1}{M} \right)~,\nonumber
\end{align}
where phase space factors $\beta_{WV}$ and the other constants are found in Appendix~\ref{app.widths}. This shows that the $WZ$ width grows with $M$, compared to the $W\gamma$ width, but this growth is tempered by $\Delta$. In the $\Delta=0$ limit, this behavior is completely absent.

\begin{figure}
    \centering
    \includegraphics[width=0.7\linewidth]{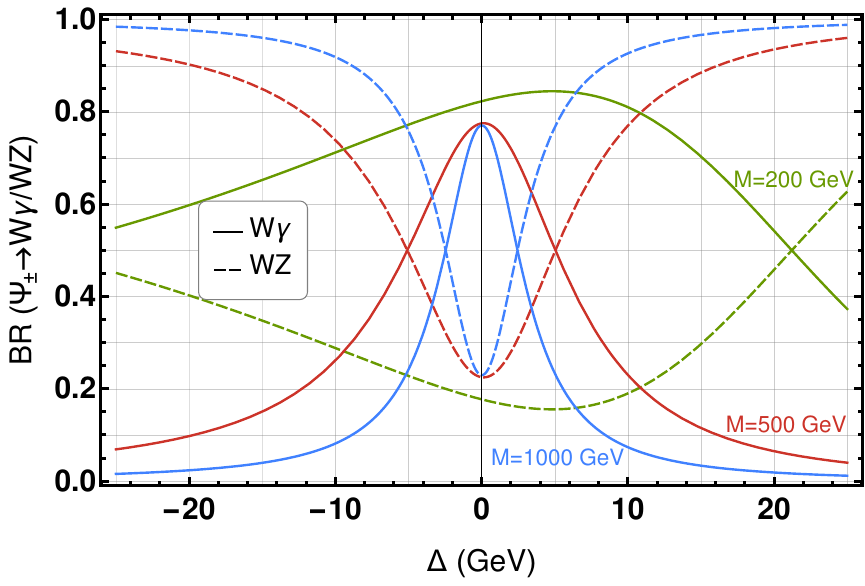}
    \caption{Branching ratios for a charged bound state $\Psi_\pm$ decaying into $W\gamma$/$WZ$ as a function of the squirk mass splitting ($\Delta$) for various total masses ($M$). The $W\gamma$ signal dominates at small $\Delta$, but quickly turns over at a few GeV. The turnover point decreases in $\Delta$ as $M$ increases.}
    \label{fig:BR-vs-Delta}
\end{figure}

In Fig.~\ref{fig:BR-vs-Delta} we plot the branching ratios of the charged bound state $\Psi_\pm$ into $W\gamma$ (solid) and $WZ$ (dashed) final states for several values of the total mass $M$ using the full formulas for the widths given in Appendix \ref{app.widths}. We see that for small $\Delta$ the branching into $W\gamma$ dominates, but also that this behavior changes as the splitting increases. The larger $M$ is, the sooner the turnover, which agrees with Eq.~\eqref{eq:widthRatio}. 

The smaller $M$ results show that there is a difference between positive and negative $\Delta$. This is due to the difference in the electric charges of the two states (in our plot we have defined positive $\Delta$ to correspond to a heavier up-type squirk). This asymmetry is significant only at low $M$. Because the branching ratio curves are essentially symmetric for $M\gtrsim 400$~GeV in the collider study that we outline in Sec.~\ref{s.collider}, we use $0<\Delta<40$~GeV ($m_{\su}\geq m_{\sd}$) as an effective range for the splitting. Considering negative $\Delta$ will change the exact results at low $M$, but should be qualitatively similar.

This new understanding of these branching fractions has significant consequences. In particular, the experimental limits that have previously been applied to natural squirks (like those of folded SUSY) must be reassessed for nondegenerate squirk masses. Since the $WZ$ final states are experimentally more challenging, the total experimental reach is expected to decrease as $\Delta$ is increased. 

To update the experimental limits we first obtain the LHC cross section to produce the $\Psi_\pm$ bound state. This was done both by convolving the partonic cross section given in Eq.~\eqref{e.sigma_production} with the proton PDFs~\cite{Martin:2009iq} and by using MadGraph~\cite{Alwall:2014hca}. We find excellent agreement between the two methods, but we use the MadGraph results in order to make a consistent comparison with the search strategy we outline in Sec.~\ref{s.collider}. This total LHC cross section was then multiplied by the appropriate bound state branching ratio ($W\gamma$ or $WZ$). Finally, we applied the branching ratio of the $W$ or $Z$ decays appropriate to the LHC search in question.

\begin{figure}[t]
    \centering
    \begin{subfigure}{0.49\textwidth}
        \includegraphics[width=\textwidth]{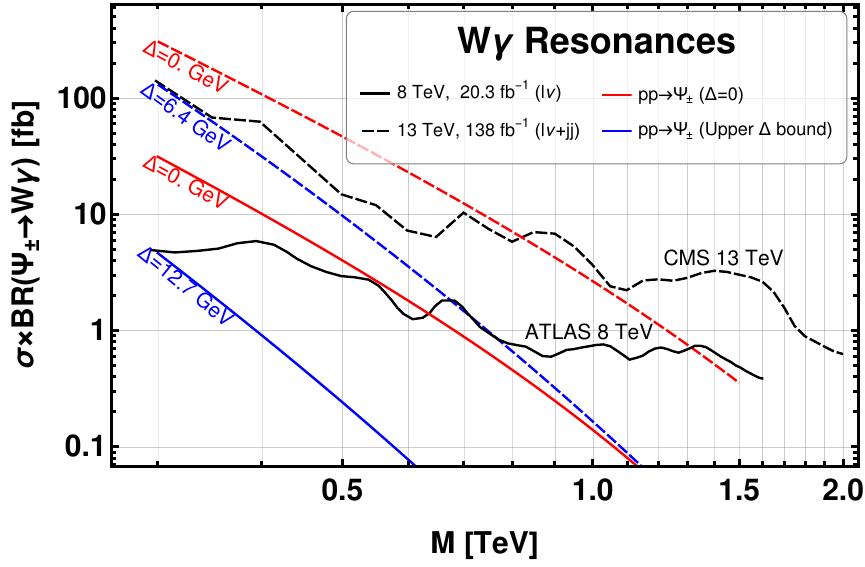}
        \caption[Image 1]{\label{fig:bounds_Wgamma}}
    \end{subfigure}
    \begin{subfigure}{0.49\textwidth}
        \includegraphics[width=\textwidth]{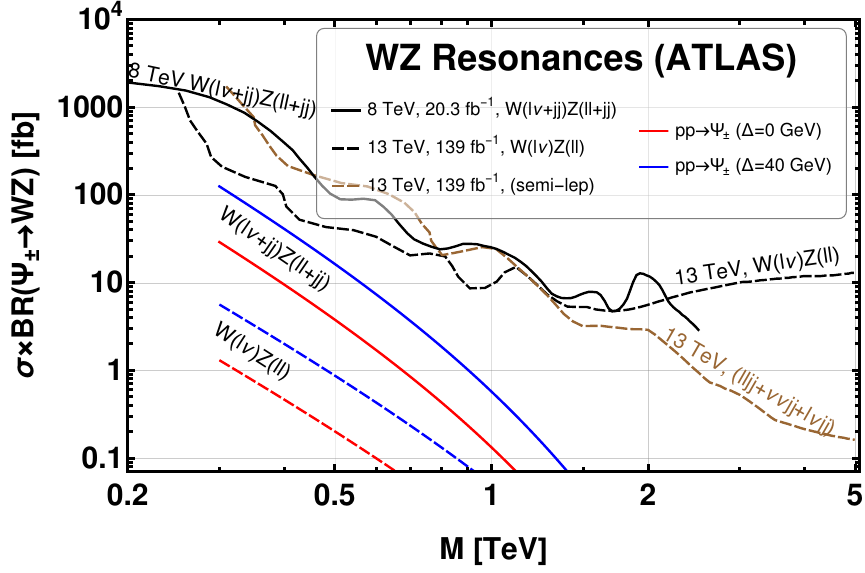}
        \caption[Image 2]{\label{fig:bounds_WZ}}
    \end{subfigure}
    \caption{Application of LHC diboson resonance searches to charged bound state $\Psi_\pm$ production and decay. (a) $W\gamma$ resonance searches lead to experimental bounds on squirk masses. See Fig.~\ref{fig:exclusion} for bounds in the $(M,\Delta)$ parameter space. (b) Current $WZ$ resonance searches provide no additional experimental bounds, even when considering squirk mass splitting up to $\Delta=40~{\rm GeV}$.}
    \label{fig:bounds}
\end{figure}

Our results are shown in Fig.~\ref{fig:bounds}. The left panel considers the LHC $W\gamma$ resonance searches~\cite{ATLAS:2014lfk,CMS:2024ndg} while the right panel uses the $WZ$ resonance searches~\cite{ATLAS:2020fry,ATLAS:2022zuc}. We plot the $\Psi_\pm$ decay signal with $\Delta=0$ (in red) for comparison with previous works. In addition, we plot the cross section at the maximum value of $\Delta$ (in blue) for which there is any experimental bound. For the $WZ$ searches we simply plot the $\Delta=40$ GeV lines for comparison. Solid lines indicate a collider energy $\sqrt{s}=8$~TeV, while dashed lines correspond to $\sqrt{s}=13$~TeV.  

The $W\gamma$ results on the left panel of Fig.~\ref{fig:bounds} illustrate the importance of $\Delta$. In the $\Delta=0$ case the 13 TeV search by CMS is probes higher masses, reaching $M\sim800$ GeV. However, for a $\Delta$ of about 10~GeV the CMS search is not sensitive to the squirk bound state. In this case the 8~TeV ATLAS search produces a much lower bound on the mass, but even this disappears by the time $\Delta=13$~GeV. 

In contrast, the $WZ$ branching fraction increases as $\Delta$ increases, as seen in the right panel of Fig.~\ref{fig:bounds}. Despite this enhancement, however, we see that the experimental searches are not sensitive enough to discover these bound states, even up to $\Delta=40$~GeV. We take this value our upper bound, because larger splitting leads to a short $\beta$-decay half-life and the charged bound state $\Psi_\pm$ becomes a neutral bound state $\Psi_0$ before it annihilates, see Sec.~\ref{ss.model_constraints}.

\begin{figure}[ht]
    \centering
    \includegraphics[width=0.7\linewidth]{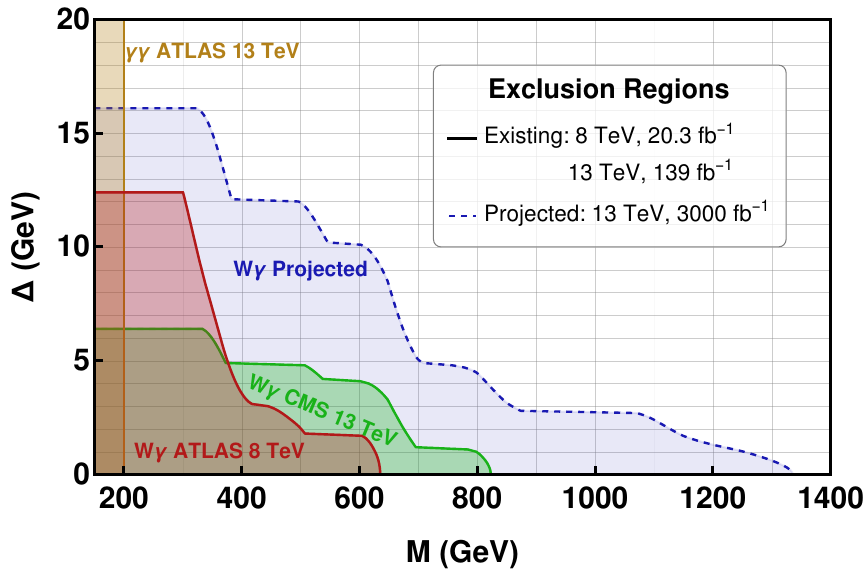}
    \caption{Direct experimental bounds on our simple squirk model as a function of the squirk mass splitting $\Delta$ and the total mass $M$. Excluded regions from completed $\gamma\gamma$ and $W\gamma$ searches are bounded by a solid line. The blue region with dashed boundary shows the projected sensitivity for $3000~{\rm fb}^{-1}$ of integrated luminosity with collider energy $\sqrt{s}=13~{\rm TeV}$.}
    \label{fig:exclusion}
\end{figure}

The direct experimental bounds on this simple squirk model are summarized in Fig.~\ref{fig:exclusion}. The bound from neutral bound states decaying to $\gamma\gamma$ holds for all mass splitting, so it appears as a vertical line in the $\Delta$-$M$ plane, with lower values of $M$ excluded. On the other hand, the bounds from charged bound state decays only apply to a limited range of $\Delta$. The figure also shows in exactly which circumstances the 13~TeV $W\gamma$ bound is stronger than the 8~TeV bound.

Perhaps the most important lesson to be taken from this plot is that there are motivated BSM states with order one electroweak charges and masses of only a few hundred GeV that are currently unconstrained by experiment. Indeed, as $M$ is the sum of the two squirk masses, currently, states with masses even as low as 100~GeV are allowed. This simple model stands a case in point that there is a great deal of unexplored BSM physics that the LHC can discover.

The immediate follow-up question is: ``What can be discovered during the high-luminosity run?'' We make a simple projection to get an idea for how the sensitivity of these search methods will change during the LHC's high-luminosity run.  Since the center-of-mass energy is the same, this is done by relating the cross section sensitivity at high luminosity $\sigma_\text{HL}$ to the current cross section sensitivities $\sigma_0$ by
\beq
\sigma_\text{HL}=\sigma_0\sqrt{\frac{\mathcal{L}_0}{\mathcal{L}_\text{HL}}}~,
\eeq
where $\mathcal{L}_0$ and $\mathcal{L}_\text{HL}$ are the luminosities associated with current searches and the high-luminosity run, respectively.

This projection is included in Fig.~\ref{fig:exclusion} as the blue region with dashed outline.  Although this projection is only a simple estimate, it gives an idea of how the sensitivity might increase over the planned LHC runs. We find that the discovery reach at $\Delta=0$ goes up considerably. The reach in $\Delta>0$ also increases along the entirety of the bound. However, there remains significant, motivated parameter space for which current search strategies are unlikely to discover these states, even for relatively low masses.

In the following section we outline a novel search strategy which often has complementary sensitivity to existing methods. By leveraging the dynamics of hidden sector glueballs we can probe regions of much larger mass splitting. We also find regions of overlapping sensitivity that offer exciting opportunities to test whether BSM signals in these channels can be associated with the squirk
scenarios discussed in this work.

\section{New Collider Strategy} \label{s.collider}

In the previous section, we determined that squirk bound states, such as arise in folded SUSY, are only beginning to be probed by the LHC.  New techniques are required for a thorough search for such states. In this section, we outline one possible approach that employs the hidden sector glueball spectrum associated with the squirks' confining gauge group. 

This hidden sector glueball can produce a displaced vertex in addition to the diboson resonance. As seen in Fig.~\ref{fig:glueballDecayLength}, only hidden sector glueballs with masses larger than about 10 GeV are likely to decay within detector subsystems with meaningful abundance, and heavier hidden sector glueballs have a higher probability to decay.  This means that there is some separation in mass between SM particles with long lifetime and our signal hidden sector glueballs. The combination of such a displaced vertex along with a diboson resonance is expected to
be background-free. Consequently, this search strategy can significantly improve the LHC's potential to discover squirks during its high luminosity run.

\subsection{Resonance + Displaced Vertex Method}\label{ss.R+DV_signals}
\begin{figure}[h]
  \centering
    \begin{fmffile}{production_with_gluon}
        \begin{fmfgraph*}(\feynwidth,\feynheight)
        \fmfleft{i1,i2}
        \fmfright{o1,o2,o3}
        \fmf{fermion}{i2,v1,i1}
        \fmf{boson,label=$W^\pm$}{v1,v2}
        \fmf{scalar}{o1,v2,v3,o3}
        \fmf{gluon,tension=0.1}{v3,o2}
        \fmfv{label=$\overline{q}_2$,label.dist=0.5,label.angle=135}{i1}
        \fmfv{label=$q_1$,label.dist=0.5,label.angle=-135}{i2}
        \fmfv{label=$\sqbar_2$,label.dist=0.5,label.angle=45}{o1}
        \fmfv{label=$\sq_1$,label.dist=0.5,label.angle=-45}{o3}
        \fmfv{label=$g'$}{o2}
        \end{fmfgraph*}
    \end{fmffile}
  \caption{Example of tree-level production of an electrically charged squirk bound state $\Psi_\pm$ and a hidden sector gluon.}
  \label{fig:productionWithGlue}
\end{figure}
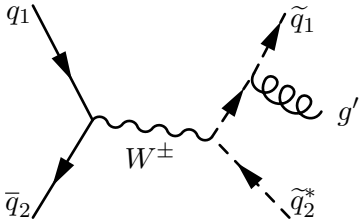

We consider an electrically charged squirk bound state that is produced with a hard hidden sector gluon,
see Fig.~\ref{fig:productionWithGlue}. The bound state obeys the usual squirk dynamics, leading to a $W\gamma$ or $WZ$ resonance on collider prompt timescales. The hidden sector gluon, however, must hadronize into one or more hidden sector glueballs. The lightest, and hence most abundantly produced, hidden sector glueball decays with long lifetime through an off-shell SM Higgs into SM states. In short, the hidden sector glueball can produce a displaced vertex within LHC detector systems. 

The cross section to produce a hidden sector gluon along with the squirks is, of course, smaller than for squirks alone. 
However, we expect the background to be significantly reduced by the presence of the squirk bound state resonance and the displaced vertex. ATLAS studied a signal of 4-7 jets and found a detector-related background of less than one event per $100~{\rm fb}^{-1}$~\cite{ATLAS:2023oti}.  Their background is primarily comprised of overlapping tracks that are misidentified as originating from the displaced decay of a heavy particle. Our signal has much less hadronic activity and can be combined with a high-$p_T$ trigger for the decay products of the energetic bosons. Another possible refinement would be to target a particular displaced decay of the hidden sector glueball. In sum, we expect that treating our search as background-free is a very good approximation.

We use MadGraph~\cite{Alwall:2014hca} to compute the cross section for the production of two squirks and a final-state hidden sector gluon. This massless hidden sector gluon must then be matched to a massive hidden sector glueball state in a way that conserves energy and momentum. To accomplish this we modify the total momentum of the squirk bound state and the gluon to allow for a nonzero hidden sector glueball mass. This is done in the CM frame by fixing the energy of the total state produced through the $W$. 

Specifically, the energy of the bound state (with mass $M$) and the massless hidden sector gluon (whose three momentum has a magnitude of $p_g$) is taken to be equal to the energy of the bound state plus a massive hidden sector glueball with three-momentum $p_G$. More concisely, we require
\beq
    \sqrt{M^2+p_g^2} + p_g = \sqrt{M^2+p_G^2} + \sqrt{m_0^2 + p_G^2}~,
\eeq
where $m_0$ is the hidden sector glueball mass. In this equation we require $p_g\geq m_0$. In other words, we require the energy of the radiated gluon to be larger than the glueball mass. Solving for $p_G$ we find
\beq \label{e.p_G}
    p_G = p_g\sqrt{
    1+\frac{m_0^2}{M^2}\left(\sqrt{1+\frac{M^2}{p_g^2}}-1\right)\left(1-\frac{m_0^2}{2M^2}\right)
    -\frac{m_0^2}{p_g^2}\left(1-\frac{m_0^2}{4M^2}\right)}~,
\eeq
which is always real valued for $p_g\geq m_0$.
MadGraph events with $p_g<m_0$ are discarded and hence the effective differential cross section is reduced as $m_0$ is increased. The momentum magnitude from Eq.~\eqref{e.p_G} is used for both the squirk bound state and the hidden sector glueball, using the CM directions for their momenta which were produced by MadGraph. The two four-momenta are then boosted back into the lab frame and the particles are evolved to their final SM decay products.

\subsubsection{Bound State Simulation}

MadGraph cannot describe the dynamics of a quirk bound state after its constituents are produced. Therefore, we simulate the subsequent dynamics directly. This simulation, however, does not carefully model the trajectory of each particle, as these are not important to our analysis. Rather, we utilize the MadGraph events as initial conditions for the bound state and extract the quantities, such as transverse momentum and pseudorapidity of the decay products, necessary to check against the experimental cuts on the final state particles.

As discussed in Sec.~\ref{sssec:ChargedBounds}, the final states produced by bound states is sensitive to the mass splitting $\Delta$ of the constituents. At the same time, a nonzero $\Delta$ introduces the possibility for the charged bound state $\Psi_\pm$ to $\beta$-decay into a neutral bound state $\Psi_0$, which cannot be discovered using the methods we are describing. Therefore, for each squirk production event we estimate the bound state's deexcitation time according to Eq.~\eqref{e.t_dE} and calculate the width for $\beta$-decay using Eq.~\eqref{eq:betaInt}. From these we calculate the probability that the bound state deexcites and annihilates before $\beta$-decay according to Eq.~\eqref{e.P_DE}. In each case we draw a random number to determine whether the event in question is to be kept (assumed to deexcite) or removed (assumed to $\beta$-decay).

The bound state deexcites by radiating gauge bosons. We assume no preferred direction to this radiation so that the angular distribution of the bound state's initial three-momenta matches the distribution of the three-momenta after deexcitation. The total three-momentum of the bound state in its ground state is taken to lie along the same direction extracted from the production event, but the magnitude is reduced to account for the energy that has been radiated.

Annihilation of the ground state particles into SM bosons provides the prompt resonance of this search strategy. In its rest frame, the bound state is replaced by two back-to-back particles of the appropriate mass. The angular distribution of these particles is taken to be flat. The vector bosons are then decayed into the particles relevant to various searches and boosted into the lab frame. In this way we can map the produced bound state constituents to the resonance searches performed at the LHC. In Appendix \ref{app.cuts} we specify the cuts used for each experimental search.

\subsubsection{Glueball Simulation}

In each simulated event a hard hidden sector gluon is radiated from a squirk. This hidden sector gluon hadronizes into a hidden sector glueball, but which state is produced? As usefully tabulated in Ref.~\cite{Batz:2023zef}, in SU(3) gauge theories there are 12 glueball states which are often labeled by their $J^{PC}$ values. The lightest state, with mass $m_0$, is $0^{++}$ and as discussed in Sec.~\ref{ss:glueDyn} this state decays into SM states through an off-shell Higgs. 

The other hidden sector glueball states are typically collider stable~\cite{Batz:2023zef} so they do not produce the displaced vertices essential to our search strategy. For hidden sector glueball masses at the higher end of what we consider, some of the heavier states may decay with a short enough lifetime to appear as a displaced vertex in LHC detectors. However, in this range, the cross section to produce a hidden sector gluon with sufficient energy to make these heavy hidden sector glueball states is quite small. For simplicity, we make the conservative assumption that only the lightest hidden sector glueball produces displaced vertices. However, we must still estimate how likely the radiated hidden sector gluon is to become $0^{++}$ as compared to the other hidden sector glueball possibilities.

While the details of hidden sector glueball hadronization are beyond the scope of this work we can make robust predictions for the relative production of hidden sector glueball states. Following Ref.~\cite{Batz:2023zef}, we estimate the relative rate of producing a hidden sector glueball with spin $J$ with a thermal distribution 
\beq
	P_J \propto (2J+1)\left(\frac{m_J}{m_0}\right)^{3/2}\exp\left[-\frac{m_J-m_0}{T_\text{had}}\right]~,
\eeq
where $m_J\geq m_0$ is the mass of the spin-$J$ hidden sector glueball and $T_\text{had}$ is a temperature that characterizes hadronization. This form shows that higher-spin hidden sector glueballs receive some enhancement. However, the largest effect is the exponential cost of having higher mass. 

This penalty on higher mass states depends crucially on $T_\text{had}$. A larger temperature leads to less suppression of the higher mass states. As argued in Ref.~\cite{Batz:2023zef} this temperature is taken to be independent of the process that produces the hidden sector gluon. That is, it is a property of the confining dynamics only. In that work a mild variation of possible temperatures were considered, all of the form
\beq
    T_\text{had} = \kappa\Lambda^{\prime}=\kappa\frac{m_0}{6.8}~,
\eeq
with $\kappa\sim1$. In this work we use $\kappa=1.04$, which is the default value employed in Ref.~\cite{Batz:2023zef}. Using this relative weighting for the production of the different hidden sector glueball species, for each production event we draw a random number to determine whether the hidden sector gluon hadronizes into a $0^{++}$ state or a heavier state. Only hidden sector glueballs with mass below the hidden sector gluon's energy are considered as possible final states.

After determining whether or not the hidden sector gluon hadronizes into a $0^{++}$ state, we determine the probability that this hidden sector glueball decays within various subsystems of the ATLAS detector using the decay width given in Eq.~\eqref{e.decay-m0} and the detector volumes described in Refs.~\cite{ATLAS:2015mlf,ATLAS:2015xit}. The probability of decay is translated into the probability for detection using the efficiency ranges, which are provided in the same references. Although there are differences between the two detectors, we expect the detection probabilities to be qualitatively similar between the two experiments.

\subsection{Results}

We parametrize our simple model in terms of $c_g$ (the effective coupling of the Higgs to the gluons of SU(3$)_{\text{QCD}'}$), $m_0$ (the mass of the lightest hidden sector glueball), $M$ (the sum of the squirk masses), and $\Delta$ (the difference between the masses). As a function of these parameters we simulate the production, evolution, and SM final states of an electrically charged squirk bound state $\Psi_\pm$ and an SU(3$)_{\text{QCD}'}$ glueball. For each simulated event we determine whether the squirk decay products pass the cuts for the resonance search and the probability of detecting a displaced hidden sector glueball decay and record the fraction of the events that satisfy these requirements as $f(c_g,m_0,M,\Delta)$. 

\begin{figure}[bht]
    \centering
    \includegraphics[width=\textwidth]{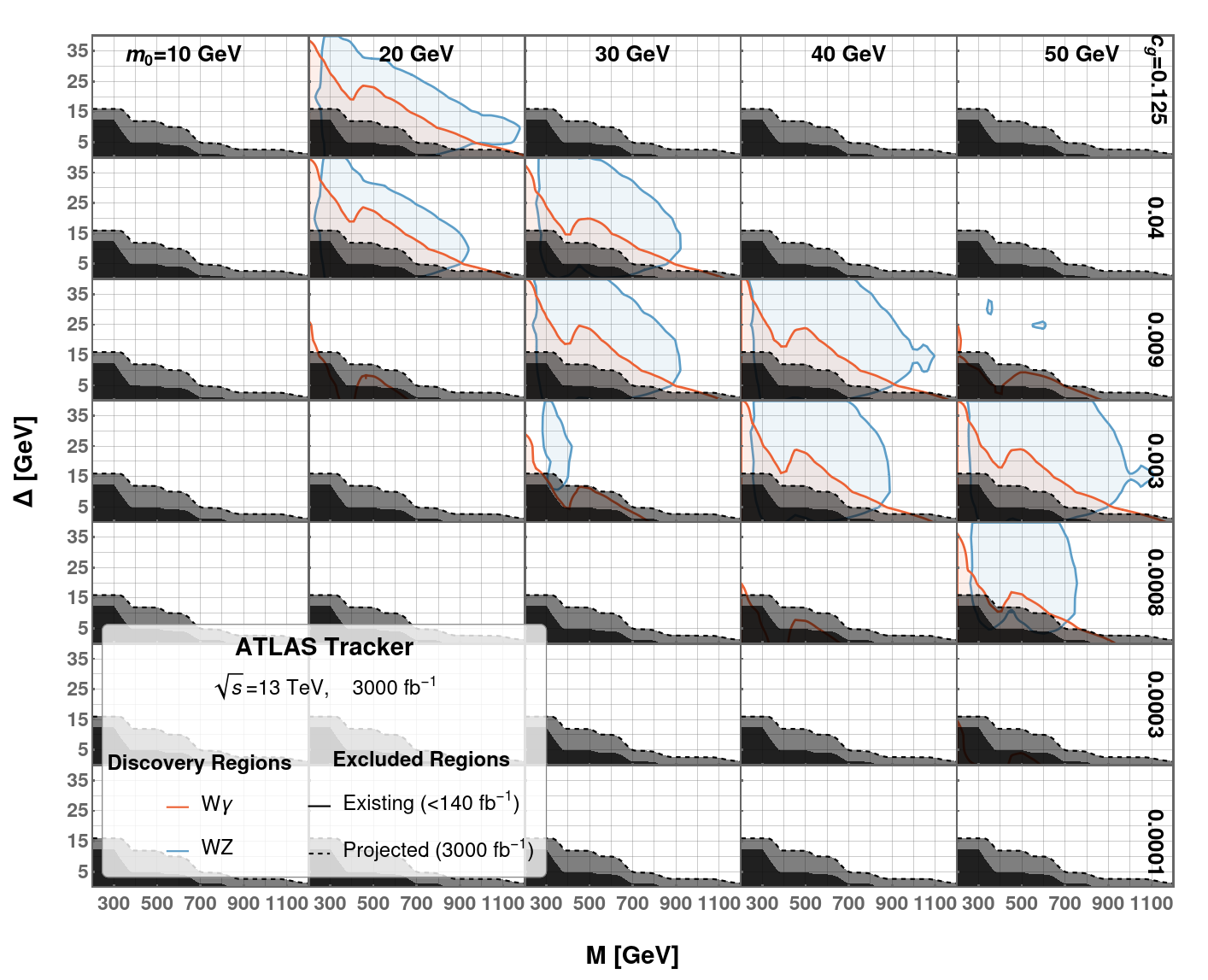}
    \caption{Exclusion and simulated discovery regions in $M$ and $\Delta$ inside the ATLAS Tracker at $\sqrt{s}=13~{\rm TeV}$ and 3000 fb$^{-1}$ for various $(c_g, m_0)$. Exclusion regions in black are due to $W\gamma$ resonance searches by ATLAS~\cite{ATLAS:2014lfk} and CMS~\cite{CMS:2024ndg}. Simulated discovery regions in orange (blue) indicate five or more events at a given $(M,\Delta)$ with a $0^{++}$ hidden sector glueball decay within the tracker and a $W^\pm\gamma$ ($W^\pm Z$) decay which passes the appropriate cuts (see Appendix \ref{app.cuts}).}
    \label{fig:events_Tracker}
\end{figure}

By using the cross sections produced by MadGraph and our calculated branching fractions we can, for a given value of integrated luminosity, translate these simulation results into the expected number of events with both a diboson resonance and a displaced vertex
\beq
    N_{WV} = \Lum\cdot\sigma(pp\to\Psi_\pm g)\cdot \text{BR}(\Psi_\pm\to WV) \cdot f(c_g,m_0,M,\Delta)~,
\eeq
where $V=\gamma,Z$. Our target scenario is the high-luminosity LHC, so we use a luminosity of $\Lum=3000$ fb$^{-1}$. Our high luminosity LHC projections use a CM energy of 13 TeV. We note that the cross sections were computed for 13 TeV collisions so comparison could be made with the cut strategies used in the 13 TeV resonance searches. However, our results should still provide a reasonable, if somewhat conservative, forecast of the discovery potential of the LHC's high-luminosity run.

\begin{figure}[t]
    \centering
    \includegraphics[width=\textwidth]{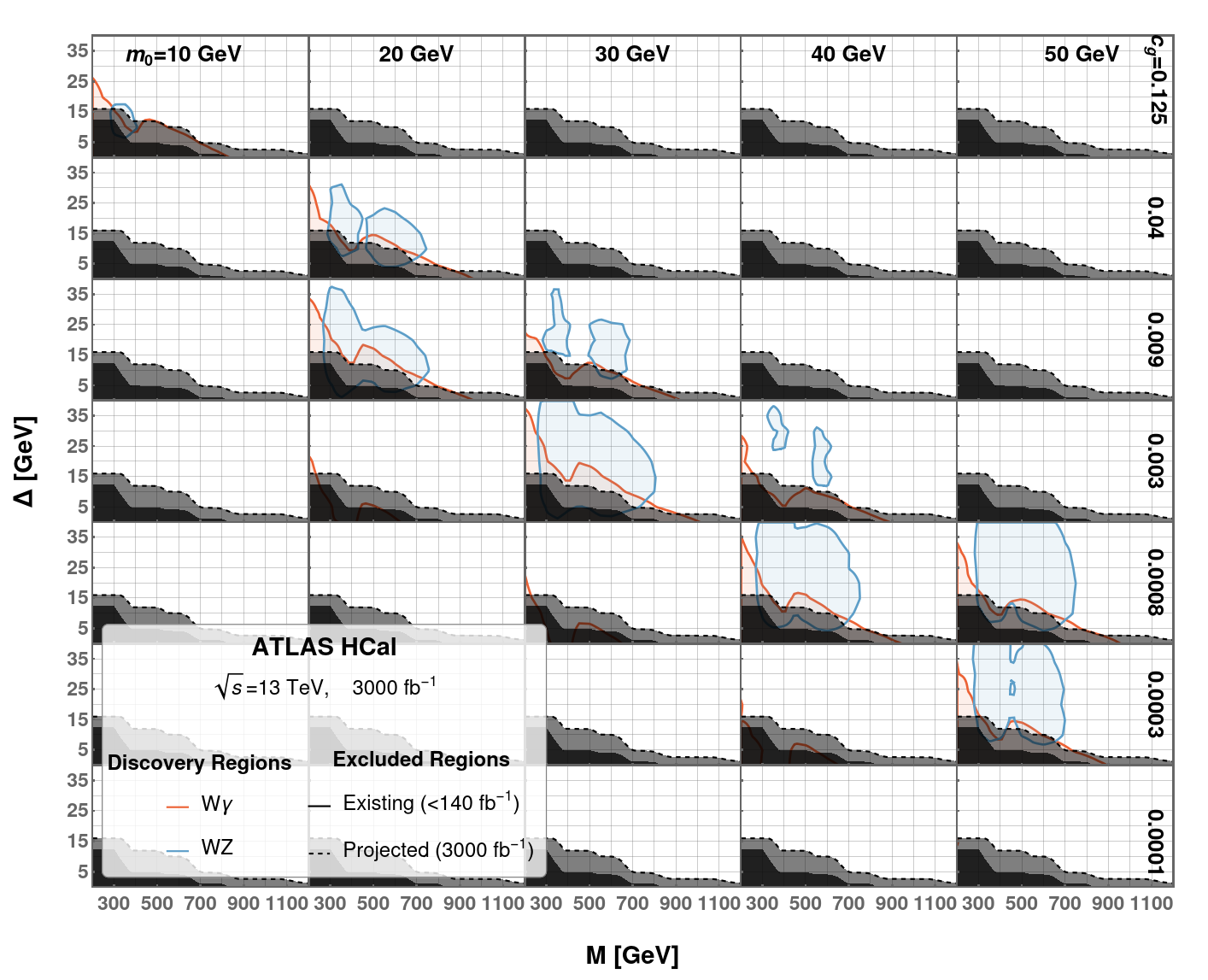}
    \caption{Exclusion and simulated discovery regions inside the ATLAS hadronic calorimeter at $\sqrt{s}=13~{\rm TeV}$ and 3000 fb$^{-1}$ given various parameter combinations $(c_g, m_0)$. See the caption to Fig.~\ref{fig:events_Tracker} for descriptions of various regions.}
    \label{fig:events_HCal}
\end{figure}
\begin{figure}[t]
    \centering
    \includegraphics[width=\textwidth]{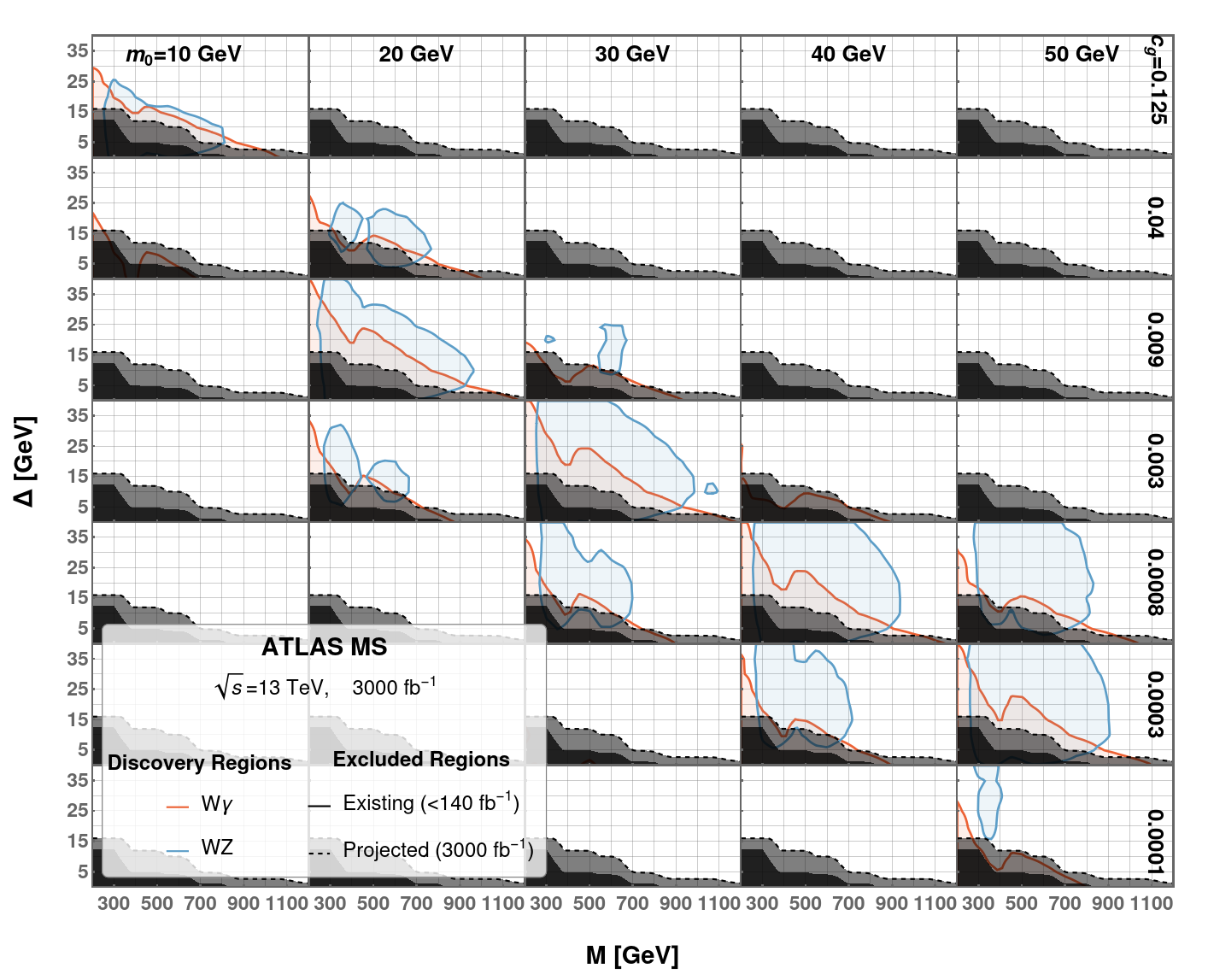}
    \caption{Exclusion and simulated discovery regions inside the ATLAS muon system at $\sqrt{s}=13~{\rm TeV}$ and 3000 fb$^{-1}$ given various parameter combinations $(c_g, m_0)$. See the caption to Fig.~\ref{fig:events_Tracker} for descriptions of various regions.}
    \label{fig:events_MS}
\end{figure}

In Figs.~\ref{fig:events_Tracker}\textendash\ref{fig:events_MS}, we display our result corresponding to displaced decays in the track, hadronic calorimeter, and muon system, respectively. We shade (in orange) regions where $N_{WV}\geq5$ as a function of $M$ and $\Delta$. The solid (dashed) curve indicates the bounds coming from $W\gamma$ ($WZ$) resonance searches. We also shade in gray the existing and projected bounds from current search methods, see Fig.~\ref{fig:exclusion}. The condition of five events is taken as a proxy for discovery, assuming essentially zero background. Each figure includes a grid of plots, scanning over several values of $c_g$ and $m_0$.

The shapes of the discovery regions are due to a combination of factors. The squirk production cross section (similar to what is shown in Fig.~\ref{fig:NeutBounds}) falls off for higher squirk masses. Also, the number of MadGraph events that lead to on-shell hidden sector glueballs goes down as the lightest hidden sector glueball mass $m_0$ goes up. In practice, however, this ends up being a small effect. In addition, the $\Psi_\pm$ branching ratio into $W\gamma$ ($WZ$) decreases (increases) as $\Delta$ grows, and this effect happens more quickly as $M$ increases, see Fig.~\ref{fig:BR-vs-Delta}. The likelihood of $\beta$-decay also increases as $\Delta$ increases. Finally, as detailed in Appendix~\ref{app.cuts}, the 13 TeV $WZ$ resonance search using fully hadronic final states employs a 500~GeV cut on the transverse momentum $p_T$ of the hardest jet along with a 200~GeV cut on the $p_T$ of the next hardest jet. The kinematics of our process are such that the two jets with highest $p_T$ typically have nearly equal values. Therefore, when the mass of the bound state reaches the TeV scale many more events begin passing this cut which enhances the signal. This leads to islands of sensitivity for $M\sim1$ TeV.

The figures show that taken together, the various detector subsystems can be used to probe a great deal of parameter space which is left untouched by current methods. The lighter hidden sector glueballs ($\sim10$ GeV) have longer decay lengths, so only the muon system begins to have sensitivity, and only for the largest values of $c_g$. As the hidden sector glueball mass increases, however, more and more values of the $c_g$ can be probed. The muon system is most sensitive to longer decay lengths, meaning smaller $c_g$, and the tracker to shorter decay lengths, meaning larger $m_0$ and $c_g$.
The hadronic calorimeter usefully interpolates between the tracker and muon system.

The figures also demonstrate how the $W\gamma$ resonance searches with a displaced vertex match the existing resonance searches. Both are most sensitive to small $\Delta$, though we do find much greater sensitivity to large $\Delta$ at smaller $M$. What is perhaps more interesting is the powerful complementarity provided by the $WZ$ resonance searches. These are most powerful for larger $\Delta$, providing sensitivity to different squirk scenarios. In particular, they illustrate the previously unappreciated point that $WZ$ resonance searches can be the dominant signal of squirk models. 

Along with the ability to probe unexplored parameter space, our results offer the potential opportunity to thoroughly investigate these types of models where the discovery regions overlap. If the high-luminosity LHC discovers evidence of BSM physics in $W\gamma$ resonances (for instance) then parallel discovery or exclusion of accompanying displaced vertices provides additional information about the nature of this new physics. These overlap regions offer the chance to over-constrain and significantly test
a natural quirk interpretation of any new signal.

\section{Conclusion\label{sec:Con}}

The LHC continues to be the discovery machine for the energy frontier. Though its searches for some types of new physics have reached the TeV scale, there remain simple, motivated extensions of the SM whose electroweakly charged particles have not yet been discovered. We have shown, using folded SUSY as an example, that the masses of these new states may still be as low as about 100 GeV. What is more, existing search strategies, even when projected to the high-luminosity run, are insufficient to discover these new states.

With this motivation we consider a simple model of two squirks that make up an electroweak doublet. These squirks can form charged bound states or neutral bound states. The charged bound states decay into pairs of electroweak bosons providing a striking resonance structure at the LHC. However, the mass difference between the squirks can make the more challenging $WZ$ final state dominate, weakening the discovery power of LHC searches.

However, like many realizations of neutral naturalness~\cite{Batell:2022pzc}, the gauge interactions that confine the squirks also produces a spectrum of hidden sector glueballs, the lightest of which decays to SM states with a long lifetime. The displaced decays of a hidden sector glueball, in association with the prompt diboson resonance, provide an opportunity to discover these types of bound states. In this article we show that this search strategy, when applied at the high-luminosity LHC, can lead to new particle discoveries that current searches would miss. Indeed, our strategy provides complementarity to existing methods, having greater power for larger mass splitting, unlike the usual approach. 

There are also regions where the present methods and our novel strategy both have sensitivity. A discovery associated with these regions presents the opportunity to stringently test whether the new physics is associated with a neutral-naturalnesslike construction. It is likely that continued effort will yield additional ways to more fully explore and examine such new physics scenarios.

One avenue which we are already pursuing is the case of neutral bound states. If these states are composed of SU$(2)_L$ singlets then the charged bound state analysis discussed in this work would not apply. Such bound states primarily decay into showers of hidden sector glueballs~\cite{Burdman:2008ek,Chacko:2015fbc}, whose modeling has recently taken significant strides forward~\cite{Curtin:2022tou,Batz:2023zef}. Currently, we are working to determine how the LHC can also effectively discover these natural squirk bound states.

\section*{Acknowledgments}
We are grateful to Pouya Asadi, Rodolfo Capdevilla, Zackaria Chacko, Ayres Freitas, Simon Knapen, and Markus Luty for valuable conversations. The research of J.F., C.T., and C.B.V. is supported in part by the National Science Foundation under Grant No. PHY-2210067. M.L. is supported by DOE Grant No. DE-SC0007914 and NSF Grant No. PHY-2112829.

\appendix

\section{Collected Formulas\label{a.Detail}}
In this appendix we record a few results useful for our phenomenological study of squirks. These are related to $\beta$-decay, the decay widths of the $\Psi_\pm$ bound state, the running of the new gauge coupling, and the cuts used in the various diboson resonance searches.

\subsection{Squirk $\beta$-decay\label{app.beta}}

A heavy squirk (with mass $m_H$) can decay to a lighter squirk (with mass $m_L$) and an off-shell $W$-boson. It is convenient to define $M=m_H+m_L$ and $\Delta=m_H-m_L$. The three-body decay width is given by
\begin{align}
    \Gamma_\beta=\frac{G_F^2\Delta^5}{6\pi^3(1+\delta)}\int_0^1 d\varepsilon \sqrt{(1-\varepsilon)(1-\varepsilon\delta^2)}\frac{4\varepsilon-\frac{(1+\delta\varepsilon)^2}{(1+\delta)^2}}{\left[1-\frac{\Delta^2}{m_W^2}\frac{(1+\delta\varepsilon)^2}{(1+\delta)^2}+\frac{\Delta^2}{m_W^2}\varepsilon \right]^2}~,\label{eq:betaInt}
\end{align}
where $\delta=\frac{\Delta}{M}$ and $m_W$ is the mass of the $W$. For $\delta\rightarrow 0$, the integral evaluates to $\frac{2}{5}$, in agreement with Ref.~\cite{Burdman:2008ek}.

\subsection{Production Cross Sections}\label{app.production}

The cross section to produce the $\su\sdbar$ bound state (which we refer to as $\Psi_\pm$) is
\beq \label{e.sigma_production}
    \sigma(u\bar{d}\to\su\sdbar) = \frac{\alpha_W^2\pi}{32s}\frac{\beta^3}{(1-m_W^2/s)^2}~,
\eeq
where $s$ is the CM energy, $\alpha_W=g^2/(4\pi)$, and
\beq
\beta=\sqrt{\left(1-\frac{M^2}{s}\right)\left(1-\frac{\Delta^2}{s}\right)}~.
\eeq
Above, $M=m_{\widetilde{u}}+m_{\widetilde{d}}$ and $\Delta=m_{\widetilde{u}}-m_{\widetilde{d}}$.
The $\sigma(d\bar{u}\to\subar\sd)$ cross section through an $s$-channel $W^-$ is identical. 

\subsection{Decay Widths\label{app.widths}}
We record the decay widths for the $s$-wave (ground state) quirk bound state of mass $M$, which we take to be approximately $m_u+m_d$. We use a formula from Ref.~\cite{Martin:2008sv} for squirk bound state decay widths into particles of type $i$ and $j$. The decay widths are given by
\beq
    \Gamma(\sq_1\sqbar_2\rightarrow ij) 
    =\frac{3}{32\pi^2(1+\delta_{ij})}\frac{|R(0)|^2}{M^2}\beta_{ij}|\overline{\mathcal{M}}_{ij}|^2~,
\eeq
with a Kronecker delta $\delta_{ij}$ to avoid double-counting final states with identical particles, $R(0)$ the radial wavefunction of the bound state at the origin, $|\overline{\mathcal{M}}_{ij}|^2$ the squared amplitude for the given process, and $\beta_{ij}$ the final state velocity given by
\beq
    \beta_{ij} 
    =\sqrt{\left( 1-\frac{m_i^2+m_j^2}{M^2} \right)^2 - 4\frac{m_i^2m_j^2}{M^4}}~.\label{eq:velBeta}
\eeq
We find
\beq \label{e.Gamma_Wgamma}
\Gamma_{W\gamma}=\frac{3\alpha_W^2 s_W^2|R(0)|^2}{2M^2}\left(1-\frac{m_W^2}{M^2} \right)\left(Q_u+Q_d-\frac{\Delta}{M}\right)^2~,
\eeq
where $Q_u$ and $Q_d$ are the electric charges of the up- and down-type squirks, respectively. In addition, $s_W=\sin\theta_W$, and $\Delta=m_u-m_d$, assuming $m_u>m_d$.

For the $WZ$ final state we find
\begin{align} \label{e.Gamma_WZ}
    \Gamma_{WZ}=&\frac{3\alpha_W^2|R(0)|^2s_W^4}{2c_W^2M^2}\beta_{WZ}\left\{ \left(Q_u+Q_d-\frac{\Delta}{M}\right)^2\right.\\
    &\left.+\frac{M^4}{8m_Z^2m_W^2}\left[\left(1-\frac{\Delta^2}{M^2} \right)\beta_{WZ}^2A_{WZ}-\left(Q_u+Q_d-\frac{\Delta}{M}\right)\left(1-\frac{m_Z^2+m_W^2}{M^2} \right) \right]^2 \right\}~,\nonumber
\end{align}
with
\begin{align}\label{e.A_WZ}
    A_{WZ}=\frac{(Q_u+Q_d)\left( 1-\frac{m_Z^2+m_W^2}{M^2}\right)+\frac{c_W^2\Delta}{s_W^2M}\left(1+\frac{m_Z^2-m_W^2}{M^2} \right)}{\left(1-\frac{m_Z^2+m_W^2}{M^2} \right)^2-\frac{\Delta^2}{M^2}\left( 1+\frac{m_Z^2-m_W^2}{M^2}\right)^2}~.
\end{align}
These results are sufficient to determine the branching fractions of the bound state into these two final states. In order to determine the numerical width, to determine the lifetime of the ground state, we must also find $|R(0)|^2$. While this cannot be generally accomplished, we make an estimate, as follows, that should be sufficient to calculate the lifetime.

The quirk ground state is in the Coulombic regime. As discussed in Refs.~\cite{Kribs:2009fy,Fok:2011yc,Harnik:2011mv}, in these circumstances we can model the quirks as experiencing the potential
\beq
V(r)=-\frac{C_2\alpha'(r)}{r}~,
\eeq
where $C_2=4/3$ is the quadratic Casimir for SU(3) gauge theories and $\alpha'(r)$ is the value of the hidden sector strong coupling evaluated at the scale $r$. The value of the ground state wavefunction at the origin is then
\beq
|\psi(0)|^2=\frac{1}{\pi a_0^3}~,
\eeq
where $a_0$ is the Bohr radius for this system and is given by
\beq
a_0^{-1}=C_2\alpha'(a_0)\frac{M^2-\Delta^2}{4M}~,
\eeq
where we have expressed the usual reduced mass in terms of $M$ and $\Delta$. Finally, the radial part of the wave function is related to the total wavefunction by a factor of the appropriate spherical harmonic $Y_{\ell m}$ function. Since the squirks decay from the ground state, with $Y_{00}=1/\sqrt{4\pi}$, we have
\beq
|R(0)|^2=\frac{|\psi(0)|^2}{|Y_{00}|^2}=4\left[ \alpha'(a_0)\frac{M^2-\Delta^2}{3M}\right]^3~.
\eeq

\subsection{Running of $\alpha'$\label{apss:runalpha}}

Here, we use the running of $\alpha'$ to find its values at a couple of significant mass scales. Following Ref.~\cite{Ellis:1996mzs} we use the formula
\beq
    \Lambda'=Q\exp\left[-\frac{1}{2b_0\alpha_s(Q)}\right]\left(\frac{b_1}{b_0}+\frac{1}{b_0\alpha_s(Q)} \right)^{b_1/(2b_0)}~,
\eeq
to define the confining scale $\Lambda'$ of the new confining gauge group. Here $Q$ is the scale at which the coupling is evaluated. Because there are no states charged under the confining SU($N$) group at low scales we have
\beq
    b_0=\frac{11N}{12\pi}~, 
    \qquad\qquad 
    b_1=\frac{17N}{22\pi}~.
\eeq
In the hidden sector glueball decay width we need the value of $\alpha'(m_0)$ at the hidden sector glueball mass. This leads to
\beq
    \Lambda'=m_0\exp\left[-\frac{1}{2b_0\alpha_s(m_0)}\right]\left(\frac{b_1}{b_0}+\frac{1}{b_0\alpha_s(m_0)} \right)^{b_1/(2b_0)}~,
\eeq
and the constraint
\beq
    6.8\exp\left( -\frac{2\pi}{11\alpha^\prime(m_0)}\right)=\left(\frac{102}{121}+\frac{4\pi}{11 \alpha^\prime(m_0)} \right)^{-\frac{51}{121}}~,
\eeq
where we have taken $m_0=6.8\Lambda'$~. This has the solution $\alpha'(m_0)\approx 0.21$ as quoted in Eq.~\eqref{e.alpha-m0}. Similarly, $\alpha'(m_Z)$ can be found by solving
\beq
    \frac{6.8 m_Z}{m_0}\exp\left( -\frac{2\pi}{11\alpha^\prime(m_Z)}\right)=\left(\frac{102}{121}+\frac{4\pi}{11 \alpha^\prime(m_Z)} \right)^{-\frac{51}{121}}~.
\eeq
For a hidden sector glueball mass of $m_0=30~{\rm GeV}$, this results in a value of $\alpha'(m_Z)\approx 0.14$.

\subsection{Glueball Decay Constant\label{app.f0}}
Extracting the glueball decay constant $f_{0^{++}}$ from lattice calculations~\cite{Chen:2005mg,Meyer:2008tr} can be nontrivial. Much of this is due to differences in notation and normalization. Both of these references report their results in terms of a value $s$.

In~\cite{Chen:2005mg}, $s=g_s^2f_{0^{++}}=4\pi\alpha_sf_{0^{++}}$ and in their conclusion they report a value of
\beq
s=15.6\pm3.2~(\text{GeV})^3~.
\eeq
However, this result is assuming that the hadronic scale parameter $r_0$ takes its SM value of $r_0^{-1}=410\pm20$ MeV. The more useful result is the dimensionless quantity $sr_0^3$ which we find to be 
\beq
sr_0^3=226\pm57~.
\eeq

These authors also determine the spectrum of glueball masses in terms of $r_0$. In Table VIII we find that the $0^{++}$ glueball mass satisfies
\beq
m_0r_0=4.16\pm0.11~.
\eeq
Because $r_0$ is common to both results we can therefore express $s$ (or $f_{0^{++}}$) in terms of the glueball mass. We find 
\beq
4\pi\alpha_sf_{0^{++}}=s=(3.14\pm0.83)m_0^3~.\label{e.f0Old}
\eeq
We note that the foundational work~\cite{Juknevich:2009gg} finds $4\pi\alpha_sf_{0^{++}}=3.06m_0^3$ which may be due to using the glueball mass
\beq
m_0r_0=4.21\pm0.11~,
\eeq
obtained in Ref.~\cite{Morningstar:1999rf}. In any case, we find it most consistent to use the same source for the glueball mass and the gluon matrix element, which leads to Eq.~\eqref{e.f0Old}. 

Reference~\cite{Meyer:2008tr} also gives results in terms of a parameter labeled $s$, but this differs from the definition in Ref.~\cite{Chen:2005mg} so we refer to it as $s'$. Specifically, $s'$ is smaller than $s$ by a factor of $16\pi^2/11$. They find
\beq
s'r_0^3=11.6\pm1.1~,
\eeq
and the lightest glueball mass
\beq
m_0r_0=3.958\pm0.047~.
\eeq
These lead to
\beq
4\pi\alpha_sf_{0^{++}}=\frac{16\pi^2}{11}s'=(2.69\pm0.27)m_0^3~. \label{e.f0New}
\eeq
We see that the two results differ, but overlap in their uncertainties. In this work we use the latter, more recent, result given in Eq.~\eqref{e.f0New}. 

\subsection{Cut Tables\label{app.cuts}}
The cuts used in Refs.~\cite{ATLAS:2023kcu,CMS:2024ndg} for $W\gamma$ resonance searches and also used in our simulation are found in Table~\ref{tab.cuts_Wgamma}. The parameters used are the transverse momentum, $p_T$, and the pseudorapidity, $\eta$, for each of the relevant particles, including the missing $p_T$ for the neutrino.

The transverse mass is defined as
\beq
    m_T = \sqrt{\left(E_T^\gamma+E_T^\ell+p_T^\text{miss}\right)^2 - \left|\vec{p}_T^{\,\gamma}+\vec{p}_T^{\,\ell}+\vec{p}_T^\text{~miss}\right|^2}~,
\eeq
where $E_T^i$ is the transverse energy of the $i$th particle given by
\beq
    E_T^i = \sqrt{m_i^2+|\vec{p}_T^{\,i}|^2}~.
\eeq
The transverse mass is used to set cuts on the transverse momentum of the photon, $p_T^\gamma$.
\begin{table}[b]
\centering
\begin{tabular}{|c|c|c|}
\hline
{\bf Quantity} & {\bf Leptonic Cuts} & {\bf Hadronic Cuts} \\
\hline
\multirow{2}{*}{$p_T$ (GeV)} & $p_T^e > 35$ & \multirow{2}{*}{$p_T^J > 200$} \\
& $p_T^\mu > 30$ & \\
\hline
$p_T^{\gamma}$ (GeV)& $0.4 \, m_T^{\ell\gamma} < p_T^{\gamma} < 0.55 \, m_T^{\ell\gamma}$ & $p_T^{\gamma} > 200$ \\
\hline
$p_T^{\text{miss}}$ (GeV)& $p_T^{\text{miss}} > 40$ & - \\
\hline
\multirow{2}{*}{$\eta^{\ell,J}$} & $\eta^e < 2.1$ and $\eta^e\not\in [1.37,1.52]$ & \multirow{2}{*}{$\eta^J < 2.0$} \\
& $\eta^\mu < 2.5$ & \\
\hline
$\eta^{\gamma}$ & \multicolumn{2}{c|}{$\eta^\gamma < 1.37$}\\
\hline
\end{tabular}
\caption{\label{tab.cuts_Wgamma} Cuts for the $W(\ell\nu)\gamma$ and $W(JJ)\gamma$ signals.}
\end{table}
A similar table for the cut strategies used in the hadronic and leptonic $WZ$ searches~\cite{ATLAS:2019nat,ATLAS:2022zuc} is seen in Table~\ref{tab:ATLAScuts1}. A new quantity is introduced
\beq
    \mathcal{R}_{p_T} \equiv \frac{{\rm min}(p_T^W,p_T^Z)}{m_{WZ}}~,
\eeq
which ensures that the transverse momenta for both the $W$ and $Z$ are larger than the invariant mass of the two vectors. 

Table~\ref{tab:ATLAScuts2} shows the cut strategies for the semileptonic $WZ$ resonance search~\cite{ATLAS:2020fry}. It is broken into the three most easily detected cases, labeled by number of leptons. For clarity, the 0-lepton case refers to when the $W$ decays hadronically and the $Z$ decays into neutrinos, the 1-lepton case refers to when the $W$ decays leptonically and the $Z$ decays hadronically, and the 2-lepton case refers to when the $W$ decays hadronically and the $Z$ decays into two visible leptons.

\begin{table}[t]
\centering
\begin{tabular}{|c|c|c|}
\hline
{\bf Quantity} & {\bf Leptonic Cuts} & {\bf Hadronic Cuts} \\
\hline
\multirow{2}{*}{$p_T$ (GeV)}& $p_T^{\ell} > 25$ & $p_T^{J_1} > 500$ (leading)\\
& $p_T^{\ell} > 27$ (trigger-matched) & $p_T^{J_2} > 200$ (subleading)\\
\hline
$p_T^{\text{miss}}$ (GeV) & $p_T^{\text{miss}} > 25$ & -\\
\hline
$E_T^{\text{miss}}$ (GeV)& $E_T^{\text{miss}} > 25$ & -\\
\hline
\multirow{2}{*}{$\eta^{\ell}$} & $\eta^e < 2.47$ and $\eta^e\not\in [1.37,1.52]$ & \multirow{2}{*}{$\eta^J < 2.0$}\\
& $\eta^\mu < 2.7$ & \\
\hline
Other & $\mathcal{R}_{p_T} > 0.35$ & -\\
\hline
\end{tabular}
\caption{\label{tab:ATLAScuts1}Cuts for the $W(\ell\nu)Z(\ell\ell)$ and $W(qq)Z(qq)$ signals.}
\end{table}

\begin{table}[t]
\centering
\begin{tabular}{|c|c|c|c|c|c|}
\hline
\multirow{2}{*}{\bf Quantity} & \multirow{2}{*}{\bf 0-lepton Cuts} & \multicolumn{2}{c|}{\bf 1-lepton Cuts} & \multicolumn{2}{c|}{\bf 2-lepton Cuts} \\
&  & merged &  resolved &  merged & resolved \\
\hline
$R$ & - & $R < 1.0$ & $R \geq 1.0$ & $R < 1.0$ & $R \geq 1.0$ \\
\hline
$p_T^{\ell}$ (GeV)& - & \multicolumn{4}{c|}{$p_T^{\ell} > 30$}\\
\hline
$p_T^{\text{miss}}$ (GeV)& $p_T^{\text{miss}} > 50$ & \multicolumn{2}{c|}{$p_T^{\text{miss}} > 25$} & \multicolumn{2}{c|}{-} \\
\hline
$E_T^{\text{miss}}$ (GeV)& $E_T^{\text{miss}} > 250$ & $E_T^{\text{miss}} > 100$ & $E_T^{\text{miss}} > 60$ & \multicolumn{2}{c|}{-} \\
\hline
\multirow{2}{*}{$p_T^{J}$ (GeV)} & \multirow{2}{*}{$p_T^{J} > 200$} & \multirow{2}{*}{$p_T^{J} > 200$} & $p_T^{J_1} > 60$ & \multirow{2}{*}{$p_T^{J} > 200$}& $p_T^{J_1} > 60$ \\
&& & $p_T^{J_2} > 45$& & $p_T^{J_2} > 30$ \\
\hline
$\eta^{J}$ & $\eta^J < 2.0$ & $\eta^J < 2.0$ & $\eta^J < 4.5$ & $\eta^J < 2.0$ & $\eta^J < 2.5$ \\
\hline
$\eta^{\ell}$ & \multirow{2}{*}{-} & \multicolumn{4}{c|}{$\eta^e < 2.47$ and $\eta^e\not\in [1.37,1.52]$} \\
& & \multicolumn{4}{c|}{$\eta^\mu < 2.5$} \\
\hline
$p_T^W$ (GeV)& - & $p_T^W > 200$ & $p_T^W > 75$ & \multicolumn{2}{c|}{-} \\
\hline
Other & \multicolumn{5}{c|}{$\mathcal{R}_{p_T} > 0.35$} \\
\hline
\end{tabular}
\caption{ATLAS cuts for the $WZ$ searches in semileptonic channels.\label{tab:ATLAScuts2}}
\end{table}

\bibliographystyle{JHEP}
\bibliography{BIB}{}

\end{document}